\documentclass[aps,prd,nofootinbib,floatfix,superscriptaddress,secnumarabic,preprintnumbers,longbibliography]{revtex4-2}
\usepackage{amsmath}
\usepackage{amsfonts}
\usepackage{graphicx}
\usepackage{subcaption}
\usepackage{xcolor}
\usepackage{commath}
\usepackage{hyperref}
\newcommand{\MSbar}{{\overline{\rm MS}}}

\begin{document}

\title{Renormalization of nonlocal gluon operators on the lattice}

\author{D.~Gavriel}
\email[]{gavriel.demetrianos@ucy.ac.cy}
\affiliation{Department of Physics, University of Cyprus, Nicosia, CY-1678, Cyprus}

\author{H.~Panagopoulos}
\email[]{panagopoulos.haris@ucy.ac.cy}
\affiliation{Department of Physics, University of Cyprus, Nicosia, CY-1678, Cyprus}

\author{G.~Spanoudes}
\email[]{spanoudes.gregoris@ucy.ac.cy}
\affiliation{Department of Physics, University of Cyprus, Nicosia, CY-1678, Cyprus}

\begin{abstract}
We study the renormalization of a complete set of gauge-invariant gluon nonlocal operators in lattice perturbation theory. We determine the mixing pattern under renormalization of these operators using symmetry arguments, which extend beyond perturbation theory. Additionally, we derive the renormalization factors of the operators within the modified Minimal Subtraction $(\rm \overline{MS})$ scheme up to one-loop. To enable a non-perturbative renormalization procedure, we investigate a suitable version of the modified regularization-invariant (${\rm RI}'$) scheme, and we calculate the conversion factors from that scheme to $\rm\overline{MS}$. The computations are performed by employing both dimensional and lattice regularizations, using the Wilson gluon action. This work is relevant to nonperturbative studies of the gluon parton distribution functions (PDFs) on the lattice.
\end{abstract}

\maketitle

\section{Introduction}

The QCD factorization theorems \cite{Collins:1989gx} allow the expression of the cross-sections for various high-energy hard processes as a convolution of a process-dependent hard scattering coefficient, computable in perturbation theory, and a parton distribution function (PDF). PDFs characterize the non-perturbative aspect of these processes, offering insights into the internal structure of hadrons. Although dependent on the renormalization scale $\mu$, PDFs are universal. These one-dimensional functions can be defined as matrix elements of specific operators in a hadron state, quantifying the longitudinal momentum fraction $x$ carried by quarks or gluons in the hadron's light-cone reference frame.

A common method for determining PDFs involves global QCD analyses of a diverse range of data collected from various high-energy experiments \cite{Ethier:2017zbq,Nocera:2014gqa,deFlorian:2009vb,Accardi:2016qay,NNPDF:2017mvq,Alekhin:2017kpj}. While these analyses provide accurate results for specific parton flavors, spin configurations, and kinematic regions, they often fall short of providing a complete picture of parton distributions. To address the limitations of experimental data in determining PDFs, it becomes necessary to study parton distributions using the theoretical framework of lattice QCD. 

In the Euclidean formulation of lattice QCD, directly determining parton distributions is not feasible due to their nature as light-cone correlation functions. Instead, they can be obtained through the Mellin moments of PDFs. In principle, it is possible to reconstruct PDFs using the Operator Product Expansion (OPE) provided a sufficient number of Mellin moments. However, in practice, only the lowest three moments have been accurately computed~\cite{LHPC:2002xzk,Dolgov:2000ca,SESAM:2002zfm,Gockeler:2002ek}. These moments are insufficient for fully reconstructing the momentum dependence of the PDFs. Obtaining precise calculations of higher moments via lattice simulations is extremely challenging due to decreases in signal-to-noise ratio and the unavoidable power-law mixing under renormalization.

Various methods are being explored for the direct calculation of PDFs using lattice QCD techniques. A recent comprehensive review can be found in Refs.~\cite{Cichy:2018mum,Ji:2020ect,Constantinou:2020pek,Cichy:2021lih,Cichy:2021ewm}. One notable approach is the quasi-distribution method, which employs the large momentum effective theory (LaMET) \cite{Ji2013,Ji2014}. Instead of directly computing light-cone correlation functions, this method calculates a Euclidean version of PDFs, called quasi-PDFs. These quasi-PDFs are defined as matrix elements of momentum-boosted hadrons coupled to gauge-invariant nonlocal operators, including a finite-length Wilson line. The resulting quasi-observable, which depends on the hadron's momentum but is independent of time, can be computed on the lattice and then renormalized nonperturbatively using an appropriate scheme. Finally, the renormalized quasi-PDF is matched to the standard PDF through a factorization formula, calculated in perturbation theory~\cite{Ma2014,Wang:2019tgg,Wang:2017qyg,Chen:2020arf,Li:2020xml}.

An alternative framework to quasi-PDFs is the Ioffe-time pseudo-distributions (pseudo-ITDs). Like quasi-PDFs, pseudo-ITDs utilize matrix elements of boosted hadrons coupled to nonlocal operators. Similar also to quasi-distributions, the pseudo-distributions approach relies on factorizing pseudo-ITDs obtained on the lattice to extract light-cone ITDs, using a matching kernel calculable in perturbative QCD. However, the necessary matching in the pseudo-distributions approach is performed at the level of ITDs in coordinate space, while in the quasi-PDFs approach, it is done in momentum space.

While quark PDFs have been extensively studied both experimentally and theoretically, the investigation of gluon PDFs has been relatively limited. However, understanding gluon PDFs is essential as gluons play a critical role in various physical measurements. Gluonic contributions make a significant impact on the proton's spin \cite{Alexandrou:2017oeh,Alexandrou:2020sml,Yang:2018bft}. Phenomenological data also suggest that gluon PDFs dominate over quark PDFs in the small-$x$ region \cite{Alekhin2014}.  Global analysis finds that accurate calculations of the gluon-dependent quantities are essential for the cross-section of Higgs boson production, heavy quarkonium and jet production \cite{Butterworth:2015oua,Sato:2019yez,Hou:2016nqm,Harland-Lang:2014zoa}, as well as for providing theoretical input to the upcoming Electron-Ion Collider \cite{Moffat2021}. In this direction, first-principle calculations of gluon PDFs using lattice QCD can significantly complement the experimental investigations. 

One complication in extracting gluon PDFs is the presence of mixing with quark flavor-singlet PDFs \cite{Collins:1984xc}. The disentanglement of the mixing will help to eliminate one of the sources of systematic uncertainties in simulations. In the case of Mellin moments of PDFs, the mixing arises during renormalization. When using quasi-PDFs or pseudo-ITDs approach, the mixing between the flavor-singlet quark and gluon PDFs should be resolved at the factorization level.

The framework for extracting the $x$-dependence of quark distributions can also be applied to gluon PDFs. This concept has been recently explored in various studies concerning quasi-PDFs \cite{Wang:2017qyg,Wang2019,Zhang:2018diq,Fan:2018dxu,Wang:2019tgg}, and the pseudo-ITDs \cite{Balitsky:2019krf,Fan:2020cpa,Fan:2021bcr,Salas-Chavira:2021wui,HadStruc:2021wmh,HadStruc:2022yaw,Khan:2022vot,Fan:2022kcb,Delmar:2023agv}. However, ab initio calculations of gluon PDFs represent a novel and relatively uncharted territory.

An important aspect of calculating PDFs from lattice QCD is the nonperturbative renormalization of quasi-PDFs. Two important features of the Wilson-line operator matrix elements in quark quasi-PDFs were revealed on the lattice in Ref. \cite{Constantinou2017}: linear divergences in addition to logarithmic divergences, and mixing among certain subsets of the original operators during renormalization. Various methods have been employed to eliminate these linear divergences, but a complete nonperturbative renormalization program was only recently developed~\cite{Alexandrou2017}. Similar effects are also expected to be present in the renormalization of nonlocal gluon operators. A recent study \cite{Zhang2018}, using the auxiliary field approach, showed that different components of nonlocal gluon operators have nontrivial renormalization patterns, making it challenging to evaluate gluon quasi-PDFs accurately. Related studies can be found in Refs.~\cite{Dorn:1981wa,Wang:2017qyg,Wang:2017eel,Wang:2019tgg,Braun:2020ymy}. 

In this work, we perform a one-loop calculation of the renormalization functions for a complete set of gluon nonlocal Wilson-line operators in both lattice and continuum regularizations. We focus on the study of mixing between these operators by considering symmetries in the standard QCD. The ultimate goal is to extract one-loop conversion factors between an appropriate regularization invariant (RI$'$) scheme, which will be applicable nonperturbatively and will respect the mixing, and the more standard minimal subtraction ($\overline{\rm MS}$) scheme.

The paper is structured as follows: In Section~\ref{sec:formulation}, we provide the setup of our calculation, including the definition of the nonlocal operators, and lattice actions used. We also provide the formulation for the renormalization of the operators and the conversion factors. Section~\ref{sec:symmetries} explores the symmetry properties of the operators. This includes their transformation under charge conjugation, parity, and time reversal, along with their symmetry properties within the rotational/octahedral point group. Section~\ref{sec:results} presents our main results at the one-loop order, for the renormalization factors of nonlocal gluon operators in the $\overline{\rm MS}$ scheme, employing dimensional and lattice regularization. We also provide appropriate renormalization conditions for an RI$'$-type scheme and we present the conversion factors between RI$'$ and $\rm \overline{MS}$ schemes. In Section~\ref{sec:conclusions}, we summarize our findings and we outline future plans. Additionally, Appendices~\ref{ap:character_table} and~\ref{ap:Feynman_parameter_integrals} contain the character table of the octahedral point group and definitions of Feynman-parameter integrals used in our results, respectively.

\section{Formulation}
\label{sec:formulation}

\subsection{Definition of operators}

The nonlocal gluon operators under study are defined in the fundamental representation as:
\begin{equation}
    O_{\mu \nu \rho \sigma} (x+ z \hat{\tau},x) \equiv 2 \ {\rm Tr} \bigg( F_{\mu \nu}(x+z \hat{\tau}) W(x+z \hat{\tau},x) F_{\rho \sigma}(x) W(x, x+z \hat{\tau})  \bigg)
    \label{eq:nonlocal_operator}
\end{equation}
where $F_{\mu \nu}$ is the gluon field strength tensor and $W(x,x+z \hat{\tau})$ denotes the straight Wilson line with length $z$. Its expression is given by the path-ordered ($\mathcal{P}$) exponential of the gauge field $A_\mu$ as follows: 
\begin{equation}
   W(x,x+z \hat{\tau}) \equiv \mathcal{P} \, \exp{\left[ i g \int_0 ^z A_\mu (x+\zeta \hat{\tau}) \, d\zeta\right]}
\end{equation}
Without loss of generality, the Wilson line is
chosen to lie along the $z$ direction: $\tau = 3$; also, the origin of the axes is placed on one of the endpoints of the operator. 

There are several relations among these operators, stemming both from their definition and from the symmetries of the QCD Lagrangian; these relations will be extensively discussed in Section~\ref{sec:symmetries}. 

Due to the antisymmetry of $F_{\mu\nu}$, for a fixed choice of the Wilson line, there are 36 nonlocal operators in total by selecting the indices of $O_{\mu \nu \rho \sigma}$ to be in any direction. However, only gluon operators that exhibit multiplicative renormalizability are appropriate for defining the gluon quasi-PDF \cite{Ma2014}. Suitable candidates for the unpolarized gluon quasi-PDF can be provided by \cite{Wang2019}:
\begin{equation}
    \label{eq:gluon_qPDF}
    {\tilde f}_{g/H}^{(n)}(x, P^z) = \mathcal{N}^{(n)} \int \frac{dz}{2\pi  x  P^z} e^{iz x P^z}  \langle H(P)| O^{(n)}(z, 0) |H(P)\rangle
\end{equation}
where $\mathcal{N}^{(n)}$ is a renormalization factor, $x$ is the longitudinal momentum fraction carried by the gluon, $P^\mu=(P^0,0,0,P^z)$ is the hadron momentum, and $H(P)$ stands for momentum-boosted hadron states. Potential candidates for the gluon operator are denoted here as $O^{(n)}(z, 0)$.

\subsection{Lattice action}

We consider a nonabelian gauge theory of $SU(N_c)$ group and $N_f$ multiplets of fermions. To simplify our calculations, we employ the Wilson plaquette gauge action for gluons:
\begin{equation}
    \hspace{-1cm}
    S=\frac{2}{g_0^2} \; \sum_{\rm plaq.} {\rm Re\,Tr\,}\{1-U_{\rm plaq.}\} + S_F
    \label{eq:action_nonlocal}
\end{equation}
where
\begin{equation}
    U_{\rm plaq.}=U_\mu(x)U_\nu(x+a\hat{\mu})U^\dagger_\mu(x+a\hat{\nu})U^\dagger_\nu(x)
\end{equation}
and $a$ stands for the lattice spacing. For simplicity, we will often omit $a$ in what follows; its presence can always be inferred by dimensional reasoning. The fermionic part of the action, $S_F$, only enters the one-loop calculation through the gluon field renormalization factor (see Sec.~\ref{sec:results}). For the sake of definiteness, we will use the clover-improved Wilson fermion action \cite{Sheikholeslami1985}; however, adapting our results to any other fermion action is trivial to one-loop order.

\medskip
A standard lattice discretization of the Wilson line in Eq.~\ref{eq:nonlocal_operator}, using gluon links $U_\tau(x)$, can be formulated as follows:
 \begin{equation}
     W(x,x+z \hat{\tau}) = \prod_{\ell=0}^{n \mp 1} U_{ \pm \tau}( x + \ell a \hat{\tau}) \, , \quad\quad n \equiv z / a
 \end{equation}
 where $U_{-\tau}(x) \equiv U_{\tau}^\dagger(x-a\hat{\tau})$ and upper (lower) signs correspond to $n > 0$ ($n < 0$). Alternative discretization methods incorporate smeared gluon links, such as stout, HYP, and Wilson flow.

Furthermore, on the lattice, $F_{\mu \nu}$ is determined by the standard clover discretization of the gluon field strength tensor, defined as follows:
\begin{equation}
    \hat{F}_{\mu \nu} \equiv -\frac{i}{8 g_0}\left(Q_{\mu \nu}-Q_{\nu \mu}\right)
\end{equation}
where $Q_{\mu \nu}$ is defined as the sum of the open plaquette loops:
\begin{equation}
\begin{aligned}
Q_{\mu \nu} & =U_\mu(x) U_\nu(x+a \hat{\mu}) U_\mu^{\dagger}(x+a \hat{\nu}) U_\nu^{\dagger}(x) \\
& +U_\nu(x) U_\mu^{\dagger}(x+a \hat{\nu}-a \hat{\mu}) U_\nu^{\dagger}(x-a \hat{\mu}) U_\mu(x-a \hat{\mu}) \\
& +U_\mu^{\dagger}(x-a \hat{\mu}) U_\nu^{\dagger}(x-a \hat{\mu}-a \hat{\nu}) U_\mu(x-a \hat{\mu}-a \hat{\nu}) U_\nu(x-a \hat{\nu}) \\
& +U_\nu^{\dagger}(x-a \hat{\nu}) U_\mu(x-a \hat{\nu}) U_\nu(x+a \hat{\mu}-a \hat{\nu}) U_\mu^{\dagger}(x) .
\end{aligned}
\end{equation}

We expect that improved gauge actions, such as the Symanzik improved action, or the implementation of stout-smeared links, will not have an impact on determining the mixing pattern under renormalization of the nonlocal operators.

\subsection{Renormalization of operators}

To study the renormalization of the nonlocal gluon operators, we choose, for convenience, to calculate the following one-particle-irreducible (1-PI) two-point bare amputated Green’s functions \footnote{For simplicity, we omit color and Lorentz indices whenever they can be understood from the context. The Green's functions under study typically depend on seven Lorentz indices: two from the external gluon fields, four from the operators, and one indicating the direction of the Wilson line.}:
\begin{equation}
    \delta^{(4)}(q+q') \ \Lambda_O (q,z) = \langle A^{a}_{\alpha} (q)\  \left(\int d^4 x \, O_{\mu \nu \rho \sigma}(x+z\hat\tau,x)\right) \ A^{b}_{\beta} (q')  \rangle_{\rm amp}
    \label{eq:Green_f_nonlocal}
\end{equation}

where operator $O_{\mu \nu \rho \sigma}$ is defined by Eq.~(\ref{eq:nonlocal_operator}), and $A^{a}_{\alpha}(q)\text{,} \ A^{b}_{\beta}(q')$ are two external gluon fields. [Use has been made of the fact that, the renormalization of $O_{\mu \nu \rho \sigma}(x+z\hat\tau,x)$ is $x$-independent due to translational invariance.]

In general, the nonlocal gluon operators may undergo mixing under renormalization. Their mixing pattern could be determined by the symmetries of the theory, as we explore in the next section. Consequently, we define the renormalization mixing matrix $Z$, which relates the bare operators to their renormalized counterparts, as follows:
\begin{equation}
  O_{(i)}^R= \sum_{j} \left(Z^{-1}\right)_{i j} O_{(j)}  
\end{equation}
Here, we use $i$ and $j$ as generic indices, to list operators within a mixing set. Note that all renormalization factors depend on the regularization $X$ (where $X=$ DR [dimensional regularization], LR [lattice regularization], etc.) and on the renormalization scheme $Y$ (where $Y=\overline{\mathrm{MS}}$, $\mathrm{RI}'$, etc.), and should thus be properly represented as $Z^{X, Y}$ unless it is clear from the context. 


The corresponding renormalized amputated Green's functions are expressed as:
\begin{equation}
    \Lambda^R_{O_{(i)}} = Z_A \sum_{j}\left(Z^{-1}\right)_{i j} \Lambda_{O_{(j)}} \,, \quad \quad A^R_{\mu} = Z_A^{-1/2} A_{\mu} 
    \label{eq:Green_f_renormalized}
\end{equation}
where $A_{\mu}$ ($A^R_{\mu}$) is the bare (renormalized) gluon field. The perturbative expansions of the operators' renormalization matrix $Z$ and the gluon field renormalization factor $Z_{A}$ are given by:
\begin{equation}
    Z_{i j}=\delta_{i j}+g^{2} z_{i j}+\mathcal{O}\left(g^{4}\right)\,, \quad \quad Z_{A}=1+g^{2} z_{A}+\mathcal{O}\left(g^{4}\right)
    \label{eq:perturb_expand_Z}
\end{equation}

To determine the mixing matrix elements $z_{ij}$ on the lattice, we perform calculations in the $\rm \overline{MS}$ scheme employing both dimensional and lattice regularization. Subsequent to the computation of the $\overline{\mathrm{MS}}$ renormalized Green's functions in DR, the process of extracting $z_{ij}^{LR, \overline{\mathrm{MS}}}$ follows from the requirement that the renormalized Green's functions be independent of the regularization:
\begin{equation}
\Lambda_{O_{(i)}}^{\rm DR, \overline{\mathrm{MS}}}=\left. \Lambda_{O_{(i)}}^{\rm LR, \overline{\mathrm{MS}}} \right|_{a \rightarrow 0}
\end{equation}
 Here, $\Lambda_{O_{(i)}}^{ \rm DR, \overline{\mathrm{MS}}}$ ($\Lambda_{O_{(i)}}^{ \rm LR, \overline{\mathrm{MS}}}$) denotes the $\overline{\mathrm{MS}}$ renormalized Green's function of operator $O_{(i)}$, computed in dimensional (lattice) regularization.
 
After replacing the right-hand side of the above equation with the expressions provided in Eqs.~(\ref{eq:Green_f_renormalized}), and ~(\ref{eq:perturb_expand_Z}), we obtain:
\begin{equation}
    \Lambda_{O_{(i)}}^{ \rm DR, \overline{\mathrm{MS}}} -\Lambda_{O_{(i)}}^{\rm LR}=g^{2}\left(z_{A}^{\rm LR, \overline{\mathrm{MS}}}-z_{ii}^{\rm LR, \overline{\mathrm{MS}}}\right) \Lambda_{O_{(i)}}^{\text{tree}} - g^{2} \sum_{j \neq i} z_{ij}^{\rm LR, \overline{\mathrm{MS}}} \Lambda_{O_{(j)}}^{\text{tree}} + \mathcal{O}\left(g^{4}\right) 
    \label{eq:diff_DR_LR}
\end{equation}
where $\Lambda_{O_{(i)}}^{\rm LR}$ denotes the bare Green's function in LR. The Green's functions appearing on the left-hand side of Eq.~(\ref{eq:diff_DR_LR}) represent the main calculations of this study. In the absence of mixing, the renormalization matrix $Z_{ij}^{\rm LR, \overline{\mathrm{MS}}}$ becomes diagonal ($z_{ij}^{\rm LR, \overline{\mathrm{MS}}}=0, \quad\text{for } i\ne j$) and thus the operators are multiplicatively renormalized.

\subsection{Conversion factors}

Apart from the commonly used $\overline{\rm MS}$ scheme, typically employed in phenomenological studies, we also adopt the modified regularization-invariant (${\rm RI}'$) scheme (see Subsection~\ref{subsec:RIprime}). Nonperturbative calculations of the renormalization factors cannot be directly performed within the $\overline{\rm MS}$ scheme since its definition is perturbative. Instead, they can be computed within a suitably defined variant of the ${\rm RI}'$ scheme, which is applicable in both nonperturbative and perturbative studies. Then, quantities that are renormalized in the ${\rm RI}'$ scheme, calculated in lattice nonperturbatively, can be converted to the $\overline{\rm MS}$ scheme through appropriate conversion factors between ${\rm RI}'$ and $\overline{\rm MS}$. These conversion factors, denoted as $\mathcal{C}^{\overline{\rm MS}, {\rm RI}'}$, can only be determined using perturbation theory and are regularization independent:
\begin{equation}
    \mathcal{C}^{\rm \overline{MS}, {\rm RI}'} \equiv \left(Z^{\rm LR, \overline{MS}}\right)^{-1} \left(Z^{\rm LR, {\rm RI}'}\right) = 
    \left(Z^{\rm DR, \overline{MS}}\right)^{-1} \left(Z^{\rm DR, {\rm RI}'}\right) 
\end{equation}

Hence, the evaluation of $\mathcal{C}^{\overline{\mathrm{MS}}, \mathrm{RI}^{'}}$ can be performed in DR, where computations are notably simpler compared to LR. Note that the conversion factors generally depend on the length of the Wilson line and the components of the ${\rm RI}'$ renormalization-scale four-vector. It is understood that, in the presence of mixing among $n$ operators, the conversion factor will be an $n\times n$ matrix.

The Green's functions in ${\rm RI}'$ can be directly converted to $\rm \overline{MS}$ through:
\begin{equation}
    \Lambda^{\rm \overline{MS}}_{O_{(i)}} = \frac{Z^{\rm LR, \overline{MS}}_{A}}{Z^{\rm LR, {\rm RI}'}_{A}} \sum_{j} \left[ \left(Z^{\rm LR, \overline{MS}}\right)^{-1} Z^{\rm LR, {\rm RI}'}\right]_{ij} \, \Lambda^{{\rm RI}'}_{O_{(j)}} =
    \frac{1}{\mathcal{C}^{\rm \overline{MS}, {\rm RI}'}_{A}} \sum_{j} \left[ \mathcal{C}^{\overline{\mathrm{MS}}, \mathrm{RI}^{'}}\right]_{ij} \, \Lambda^{{\rm RI}'}_{O_{(j)}}
    \label{eq:Greens_function_MSbar_to_RIprime}
\end{equation}
where the value of gluon field conversion factor $\mathcal{C}^{\rm \overline{MS}, {\rm RI}'}_{A} \equiv Z^{\rm LR, {\rm RI}'}_{A}/Z^{\rm LR, \overline{MS}}_{A} = Z^{\rm DR, {\rm RI}'}_{A}/Z^{\rm DR, \overline{MS}}_{A}$ is given by~\cite{Gracey2003}:
\begin{equation}
    \mathcal{C}^{\rm \overline{MS}, {\rm RI}'}_{A} = 1 + \frac{g^2}{16 \pi^2} \frac{\left(97 + 18 (1 - \beta) + 9 (1 - \beta)^2 \right) N_c - 40 N_f}{36} + \mathcal{O}(g^4)
\end{equation}
where $\beta$ is the standard gauge parameter: $\beta = 0 (1)$ corresponds to the Feynman (Landau) gauge.

In nonperturbative investigations of Green's functions using physical hadron states through lattice simulations, the normalization of external states is conducted without involving gluon field renormalization $Z_{A}$. Consequently, the only required conversion factor in this case is $\mathcal{C}^{\overline{\mathrm{MS}}, \mathrm{RI}^{\prime}}$.

\section{Symmetry properties}
\label{sec:symmetries}

In this Section, we make use of all available symmetries [including space-time symmetries and local BRST (Becchi-Rouet-Stora-Tyutin) invariance] to pinpoint the possible mixing sets among nonlocal gluon operators. The first observation to be made is that all mixing operators will necessarily be of the same form as Eq.~(\ref{eq:nonlocal_operator}), possibly with different values for the Lorentz indices $\mu,\,\nu,\,\rho,\,\sigma.$ This stems from the following arguments:
\begin{itemize}
\item Wilson lines renormalize multiplicatively~\cite{Dorn:1986dt} (see also Ref.~\cite{Dotsenko:1979wb} for smooth closed Wilson loops and Ref.~\cite{Brandt:1981kf} for Wilson loops involving singular points). 
\item There can be no mixing with nonlocal fermion operators, i.e. with operators having the generic form: 
\begin{equation}
    \bar\Psi(x+z \hat{\tau}) \,\Gamma \, W(x+z \hat{\tau},x) \,\Psi(x)    \label{eq:fermion_operator}
\end{equation}
where $\Psi$ generally stands for a fermion field, possibly with one or more covariant derivatives, and $\Gamma$ is a Dirac $\gamma$-matrix (or product thereof). The reason for this absence of mixing is that $\Psi$ transforms under the fundamental, rather than the adjoint, representation of the global gauge group.\footnote{A more complicated alternative, in which $\Psi$ could stand for a product of two fermion fields (and thus could transform under the adjoint representation) would lead to an operator of higher dimensionality and thus would be excluded from mixing.}
\item In principle, mixing with higher dimensional operators multiplied by the appropriate power of the lattice spacing can arise. Such kind of mixing vanishes when taking the continuum limit in the simpler case of local operators, and thus, it is typically disregarded. However, the elimination of such mixing is not obvious in the case of nonlocal operators, where power divergences $\mathcal{O} (1/a^n), \ n \in \mathbb{Z}^+$ are present, and thus, $\mathcal{O} (a^n)$ contributions in the bare Green's functions can lead to $\mathcal{O} (a^0)$ effects in the renormalized Green's functions beyond one loop. One alternative way to suppress these unwanted effects, beyond the inclusion of such higher dimensional operators in the mixing sets, is to subtract artifacts from the bare Green's functions calculated in lattice perturbation theory. This method has been successfully applied by our group in the renormalization of local quark bilinear operators~\cite{Constantinou:2009tr,Constantinou:2013ada,Alexandrou:2015sea} and of nonlocal quark Wilson-line operators~\cite{Constantinou:2022aij}. This can be a natural extension of the present calculation. 
\item As in the case of local operators, there could, {\it a priori}, exist mixing with non-gauge invariant operators, in particular\cite{Joglekar:1975nu,Collins:1984xc}: BRST variations of other operators [Class A]; operators which vanish by the equations of motion [Class B]; and \textit{finite} mixing with any other operator having the same symmetry properties [Class C]. However, it can be verified by inspection that substitution of the field strength tensor $F_{\mu\nu}$, on either side of the Wilson line, by any combination of elementary fields, would violate one or more of the symmetries, first and foremost the local BRST symmetry.
\end{itemize}
Thus, in what follows we will investigate the mixing set, exclusively among operators shown in Eq.~(\ref{eq:nonlocal_operator})

In lattice QCD, the action remains invariant under discrete transformations of charge conjugation ($\mathcal{C}$), parity ($\mathcal{P}$), and time reversal ($\mathcal{T}$) \cite{Rothe:1992nt,Gattringer:2010zz}. In what follows, we present the analysis of the symmetry properties concerning the nonlocal gluon operators under $\mathcal{C}$, $\mathcal{P}$, $\mathcal{T}$ transformations, and transformations under the discrete rotational group. Since we consider the Wilson line direction as special, we study the residual three-dimensional rotational symmetry (or the discrete rotational octahedral symmetry on the lattice). The importance of this study lies in the fact that if two operators undergo different transformations, symmetries act as a safeguard, preventing them from mixing with each other under renormalization across all orders of perturbation theory. Conversely, operators lacking protection from symmetries are generally prone to mixing.

\subsection{\texorpdfstring{$\mathcal{C}$}{TEXT}, \texorpdfstring{$\mathcal{P}$}{TEXT}, \texorpdfstring{$\mathcal{T}$}{TEXT} transformations}

First, let us review the transformations of fields under $\mathcal{C}$, $\mathcal{P}$, $\mathcal{T}$ symmetries. Since the operators under study (Eq.~(\ref{eq:nonlocal_operator})) are made out of the gluon field-strength tensor and the Wilson line, we only need to consider the transformations of links, $U_\mu$. We work in Euclidean spacetime with coordinates $(x, y, z, t ) = (1, 2, 3, 4)$ throughout this paper. 

Charge conjugation $\mathcal{C}$ acts on lattice links as:
\begin{equation}
     U_\mu (x) \xrightarrow[]{\mathcal{C}} U_{\mu} (x) ^* = (U^\dagger _\mu (x))^\top
     \label{eq:charge_conj}
\end{equation}

Since there is no distinction between time and space in the Euclidean formulation, the parity transformation, denoted as $\mathcal{P}_\mu$ with $\mu \in \{1, 2, 3, 4\}$, can be defined in any direction \cite{Gattringer:2010zz}.
\begin{equation}
\begin{split}
    U_\mu (x) &\xrightarrow[]{\mathcal{P}_\mu} U_\mu \left(\mathcal{\mathbb{P}}_\mu (x)\right) \\
    U_{\nu} (x) &\xrightarrow[]{\mathcal{P}_\mu} U_{\nu}^{\dagger} \left(\mathcal{\mathbb{P}}_\mu (x) - \hat{\nu}\right)\,, \quad \nu \ne \mu
    \label{eq:parity_revrs}
\end{split}
\end{equation}
where $\mathcal{\mathbb{P}}_\mu (x)$ is the vector $x$ with sign flipped except for the $\mu$-direction.

Analogously, for any direction in Euclidean space one may define a time reversal transformation, denoted as $\mathcal{T}_\mu$:
\begin{equation}
\begin{split}
    U_\mu (x) &\xrightarrow[]{\mathcal{T}_\mu} U^\dagger _\mu (\mathcal{\mathbb{T}}_\mu (x)-\hat{\mu}) \\
    U_{\nu} (x) &\xrightarrow[]{\mathcal{T}_\mu} U_{\nu} (\mathcal{\mathbb{T}}_\mu (x))\,, \quad \nu \ne \mu
    \label{eq:time_revers}
\end{split}
\end{equation}
where $\mathcal{\mathbb{T}}_\mu (x)$ is the vector $x$ with sign flipped in the $\mu$-direction.

Utilizing the link transformations, we can construct the transformations of the gluon field strength tensor. For charge conjugation, we find that,
\begin{align}
     F_{\mu \nu} (x) \xrightarrow[]{C} -F_{\mu \nu}(x)^\top
\end{align}
By employing the above transformation relations and the cyclic property of traces, it can be shown that under charge conjugation the operators in Eq.~(\ref{eq:nonlocal_operator}) remain invariant.

Under parity transformations, the gluon field-strength tensor transforms as:
\begin{equation}
\begin{split}
     F_{\mu \nu} (x) &\xrightarrow[]{\mathcal{P}_\mu} -F_{\mu \nu}(\mathbb{P}_\mu (x)) \\
     &\xrightarrow[]{\mathcal{P}_\nu} -F_{\mu \nu}(\mathbb{P}_\nu (x)) \\
     &\xrightarrow[]{\mathcal{P}_{\rho}} F_{\mu \nu}(\mathbb{P}_\rho (x))\,, \quad \mu\ne\rho\ne\nu
     \label{eq:parity_transformation}
\end{split}
\end{equation}

Finally, the transformation of the gluon field strength tensor  under time reversal is as follows:
\begin{equation}
\begin{split}
     F_{\mu \nu} (x) &\xrightarrow[]{\mathcal{T}_\mu} -F_{\mu \nu}(\mathbb{T}_\mu (x)) \\
     &\xrightarrow[]{\mathcal{T}_\nu} -F_{\mu \nu}(\mathbb{T}_\nu (x)) \\
     &\xrightarrow[]{\mathcal{T}_{\rho}} F_{\mu \nu}(\mathbb{T}_\rho (x))\,, \quad \mu\ne\rho\ne\nu
\end{split}
\label{eq:time_transformation}
\end{equation}

Given that some of these transformations alter the sign of $z$, it is useful to consider the translation invariance of the Lagrangian, which imposes: 
 \begin{equation}
    O_{\mu \nu \rho \sigma} (-z,0) \rightarrow O_{\mu \nu \rho \sigma} (0,z)
    \label{eq:translation_property}
\end{equation} 
and the cyclic permutations on the trace of the operators in Eq.~(\ref{eq:nonlocal_operator}):
\begin{equation}
    O_{\mu \nu \rho \sigma} (z,0) = O_{\rho \sigma \mu \nu } (0,z)
    \label{eq:cyclic_property}
\end{equation} 
Taking advantage of Eqs.~(\ref{eq:translation_property}), and ~(\ref{eq:cyclic_property}), it is convenient to perform a change of basis in the form of,
\begin{equation}
    O_{\mu \nu \rho \sigma}^{\pm} (z,0) 
    = \frac{1}{2} \left( O_{\mu \nu \rho \sigma} (z,0) \pm O_{\rho \sigma \mu \nu } (z,0) \right)
\end{equation}
where now these operators are eigenstates of parity transformations (performed with respect to the midpoint of the operators) with eigenvalues of $+1$ (even, $\text{E}$) or $-1$ (odd, $\text{O}$). Note that $O_{\mu \nu \rho \sigma}^{-} (z,0)$ vanishes when $(\mu,\nu) = (\rho,\sigma)$. This way allows us to classify the 36 operators into several categories, each demonstrating distinct transformations under parity, as illustrated in Table ~\ref{tb:parity_groups}. 

\begin{table}[ht]
\centering
\begin{tabular}{c|c|c|c|c|c|}
Operators & $\mathcal{P}_{1}$ & $\mathcal{P}_{2}$ & $\mathcal{P}_{3}$ & $\mathcal{P}_{4}$ \\
\hline

$\begin{array}[c]{@{}l@{}l@{}} O_{3131}^+ , O_{3232}^+ , O_{3434}^+ \\
O_{1212}^+ , O_{1414}^+ , O_{2424}^+ \end{array}$ & $\text{E}$ & $\text{E}$ & $\text{E}$ & $\text{E}$ \\[12pt]

$\begin{array}[c]{@{}l@{}} O_{3132}^+ , O_{4142}^+ \end{array}$ & $\text{E}$ & $\text{E}$ & $\text{E}$ & $\text{O}$ \\
$\begin{array}[c]{@{}l@{}} O_{3134}^+ , O_{2124}^+ \end{array}$ & $\text{O}$ & $\text{E}$ & $\text{E}$ & $\text{O}$ \\
$\begin{array}[c]{@{}l@{}} O_{3234}^+ , O_{1214}^+ \end{array}$ & $\text{E}$ & $\text{O}$ & $\text{E}$ & $\text{O}$ \\[5pt]
$\begin{array}[c]{@{}l@{}} O_{3132}^- , O_{4142}^- \end{array}$ & $\text{O}$ & $\text{O}$ & $\text{E}$ & $\text{E}$ \\
$\begin{array}[c]{@{}l@{}} O_{3134}^- , O_{2124}^- \end{array}$ & $\text{E}$ & $\text{O}$ & $\text{E}$ & $\text{E}$ \\
$\begin{array}[c]{@{}l@{}} O_{3234}^- , O_{1214}^- \end{array}$ & $\text{O}$ & $\text{E}$ & $\text{E}$ & $\text{E}$ \\[5pt]

$\begin{array}[c]{@{}l@{}} O_{3212}^+ , O_{3414}^+ \end{array}$ & $\text{O}$ & $\text{E}$ & $\text{O}$ & $\text{E}$ \\
$\begin{array}[c]{@{}l@{}} O_{3121}^+ , O_{3424}^+ \end{array}$ & $\text{E}$ & $\text{O}$ & $\text{O}$ & $\text{E}$ \\
$\begin{array}[c]{@{}l@{}} O_{3141}^+ , O_{3242}^+ \end{array}$ & $\text{E}$ & $\text{E}$ & $\text{O}$ & $\text{O}$ \\[5pt]
$\begin{array}[c]{@{}l@{}} O_{3212}^- , O_{3414}^- \end{array}$ & $\text{E}$ & $\text{O}$ & $\text{O}$ & $\text{O}$ \\
$\begin{array}[c]{@{}l@{}} O_{3121}^- , O_{3424}^- \end{array}$ & $\text{O}$ & $\text{E}$ & $\text{O}$ & $\text{O}$ \\
$\begin{array}[c]{@{}l@{}} O_{3141}^- , O_{3242}^- \end{array}$ & $\text{O}$ & $\text{O}$ & $\text{O}$ & $\text{E}$ \\[5pt]

$\begin{array}[c]{@{}l@{}l@{}} O_{3124}^+ , O_{3241}^+ , O_{3412}^+ \end{array}$ & $\text{O}$ & $\text{O}$ & $\text{O}$ & $\text{O}$ \\
$\begin{array}[c]{@{}l@{}l@{}} O_{3124}^- , O_{3241}^- , O_{3412}^- \end{array}$ & $\text{E}$ & $\text{E}$ & $\text{O}$ & $\text{E}$ 
\end{tabular}

\caption{Categories of operators exhibiting different parity transformations. The arguments of the operators are omitted.}
\label{tb:parity_groups}
\end{table}

Thus, operators belonging to different categories cannot mix among themselves. The mixing pattern will be further reduced in the following subsection, by taking into account octahedral symmetry. Given that time reversal transformations are merely a composition of 3 parity transformations (and vice versa), they provide no
further information on the mixing pattern. 

\subsection{Rotational octahedral point group}

The rotational octahedral point group refers to a symmetry group that describes the discrete rotational symmetry of an octahedron or a cube. This group consists of 24 elements, corresponding to rotations by various angles with respect to different axes. It possesses five irreducible representations, including two 1-dimensional representations denoted as $A_1$ and $A_2$, one 2-dimensional representation labeled as $E$, and two 3-dimensional representations labeled as $T_1$ and $T_2$. The character table can be found in Appendix~\ref{ap:character_table}.

Taking into account the classification of operators in Table~\ref{tb:parity_groups}, we can explore whether they share the same irreducible representations. Let us start with the operator triplet $O_{3131}^+, O_{3232}^+, O_{3434}^+$: it supports a 3-dimensional reducible representation, which can be decomposed into a one-dimensional representation ($A_{1}$) and a two-dimensional representation ($E$). Various choices for the basis elements of $E$ are possible, for example:
$$\begin{tabular}{lc}
    $A_1:$ & $O_{3131}^+ +O_{3232}^+ +O_{3434}^+$ \\
    \\
    $E:$  & $\left(\begin{array}{c}
            2 O_{3434}^+ -O_{3131}^+ -O_{3232}^+ \\
            O_{3131}^+ -O_{3232}^+
            \end{array}\right)$
\end{tabular}$$

Similar reasoning can be applied to the operators $O_{1212}^+$, $O_{1414}^+$, and $O_{2424}^+$. In this case, we can identify the following operators, supporting irreducible representations:
$$\begin{tabular}{lc}
    $A_1:$ & $O_{1212}^+ +O_{1414}^+ +O_{2424}^+$ \\
    \\
    $E:$  & $\left(\begin{array}{c}
            2 O_{1212}^+ -O_{1414}^+ -O_{2424}^+ \\
            O_{1414}^+ -O_{2424}^+
            \end{array}\right)$
\end{tabular}$$


Proceeding analogously, we can draw conclusions about the rest of the categories of Table~\ref{tb:parity_groups}. For the second and third set of categories, we can identify operators supporting the three-dimensional irreducible representations $T_1$ and $T_2$:
$$\begin{tabular}{ccccccccc}
    $T_1:$ &
    $\begin{pmatrix}
    O_{3132}^- \\
    O_{3431}^- \\
    O_{3234}^-
    \end{pmatrix}$
    & ,
    &
    $\begin{pmatrix}
    O_{4142}^- \\
    O_{2421}^- \\
    O_{1214}^-
    \end{pmatrix}$
    &
    \quad\quad\quad\quad & 
    $T_2:$ &
    $\begin{pmatrix}
    O_{3132}^+ \\
    O_{3431}^+ \\
    O_{3234}^+
    \end{pmatrix}$
    & ,
    &
    $\begin{pmatrix}
    O_{4142}^+ \\
    O_{2421}^+ \\
    O_{1214}^+
    \end{pmatrix}$
\end{tabular}$$


Concerning the remaining categories, the analysis becomes somewhat more intricate; we can construct linear combinations of these operators that support the following irreducible representations:
$$\begin{tabular}{lccc}
    $T_1:$ & $\left(\begin{array}{c}
            O_{3212}^+ + O_{3414}^+ \\
            O_{3121}^+ + O_{3424}^+ \\
            O_{3141}^+ + O_{3242}^+
            \end{array}\right)$ & , & $\left(\begin{array}{c}
            O_{3212}^- + O_{3414}^- \\
            O_{3121}^- + O_{3424}^- \\
            O_{3141}^- + O_{3242}^-
            \end{array}\right)$ \\
            \\
    $T_2:$ & $\left(\begin{array}{c}
            O_{3212}^+ - O_{3414}^+ \\
            O_{3121}^+ - O_{3424}^+ \\
            O_{3141}^+ - O_{3242}^+
            \end{array}\right)$ & , & $\left(\begin{array}{c}
            O_{3212}^- - O_{3414}^- \\
            O_{3121}^- - O_{3424}^- \\
            O_{3141}^- - O_{3242}^-
            \end{array}\right)$ \\
            \\
    $A_1:$ & $O_{3124}^+ +O_{3241}^+ +O_{3412}^+$ & , & $O_{3124}^- +O_{3241}^- +O_{3412}^-$\\\\
    $E:$  & $\left(\begin{array}{c}
            2 O_{3412}^+ -O_{3241}^+ -O_{3124}^+ \\
            O_{3124}^+ -O_{3241}^+
            \end{array}\right)$ & , & $\left(\begin{array}{c}
            2 O_{3412}^- -O_{3241}^- -O_{3124}^- \\
            O_{3124}^- -O_{3241}^-
            \end{array}\right)$ 
\end{tabular}$$


Combining our findings from the octahedral point group and parity transformations, we arrange the 36 operators into 16 groups, as shown in Table~\ref{tb:parity_rotation_symmetries}. We notice that the operators in groups $\{1,\ 2\}$ have exactly the same behavior under parity transformations and the octahedral group: consequently, they have the potential to mix under renormalization. The same conclusion applies to the operators in groups $\{3,\ 4\}$, $\{5,\ 6\}$, $\{7,\ 8\}$. By the same arguments, operators in
groups 9-16 cannot possibly mix; thus, quantum corrections result in a mere multiplicative
renormalization for these operators. Finally, we note that, in groups containing multiplets (doublets or triplets) the renormalization and mixing coefficients are the same for each component of the multiplet.

\smallskip
We emphasize that all the above findings, being based on symmetry properties alone, are valid beyond perturbation theory. Thus, by making use of the operators of Table~\ref{tb:parity_rotation_symmetries} in numerical simulations, one can avoid unnecessary contamination from spurious mixing contributions.

\smallskip
It is worth mentioning that the same mixing pattern will be observed in the continuum, where octahedral symmetry is replaced by $O(3)$ symmetry. This is because every mixing pair contains one operator with at least one index along the $z$-axis and one operator with no such index; such operators cannot be related via a continuum transformation, and thus they can still mix, just as on the lattice. However, in the continuum some of the $Z$ factors of different groups will coincide; this is related to the fact that the $E$ and $T_2$ representations of the cubic group combine into the spin-2 representation of the $O(3)$ group, and therefore the corresponding renormalization factors must be equal. In particular:
\begin{align}
\begin{split}
    Z^{\rm DR, \overline{MS}}_{3\,3} &= Z^{\rm DR, \overline{MS}}_{7\,7}, \quad \quad Z^{\rm DR, \overline{MS}}_{3\,4} = Z^{\rm DR, \overline{MS}}_{7\,8}, \quad \quad Z^{\rm DR, \overline{MS}}_{4\,3} = Z^{\rm DR, \overline{MS}}_{8\,7}, \\
    Z^{\rm DR, \overline{MS}}_{4\,4} &= Z^{\rm DR, \overline{MS}}_{8\,8}, \quad \quad 
    Z^{\rm DR, \overline{MS}}_{11\,11} = Z^{\rm DR, \overline{MS}}_{15\,15}, \quad \quad Z^{\rm DR, \overline{MS}}_{12\,12} = Z^{\rm DR, \overline{MS}}_{16\,16}
    \label{eq:rotations_continuum_Z}
\end{split}
\end{align}

\newpage

\begin{table}[h!]
\centering
\begin{tabular}{c|c|c|c|c|c|c|}
Group & Operators & $\mathcal{P}_{1}$ & $\mathcal{P}_{2}$ & $\mathcal{P}_{3}$ & $\mathcal{P}_{4}$ & Irreducible Repr. \\[2pt] 
\hline
1 & $O_{3131}^+ +O_{3232}^+ +O_{3434}^+$ & $\text{E}$ & $\text{E}$ & $\text{E}$ & $\text{E}$ & $A_1$ \\[5pt]
2 & $O_{1212}^+ +O_{1414}^+ +O_{2424}^+$  & $\text{E}$ & $\text{E}$ & $\text{E}$ & $\text{E}$ & $A_1$ \\[5pt]
3 & $\left(\begin{array}{c}
            2 O_{3434}^+ -O_{3131}^+ -O_{3232}^+ \\
            O_{3131}^+ -O_{3232}^+
            \end{array}\right)$ & $\text{E}$ & $\text{E}$ & $\text{E}$ & $\text{E}$ & $E$ \\[15pt]
4 & $\left(\begin{array}{c}
            2 O_{1212}^+ -O_{1414}^+ -O_{2424}^+ \\
            O_{1414}^+ -O_{2424}^+
            \end{array}\right)$ & $\text{E}$ & $\text{E}$ & $\text{E}$ & $\text{E}$ & $E$ \\[15pt] 
5 & $\left(\begin{array}{c}
            O_{3132}^- \\
            O_{3431}^- \\
            O_{3234}^-
            \end{array}\right)$ & $\begin{array}{c}
            \text{E} \\
            \text{E} \\
            \text{O}
            \end{array}$ & $\begin{array}{c}
            \text{E} \\
            \text{O} \\
            \text{E}
            \end{array}$ & $\begin{array}{c}
            \text{E} \\
            \text{E} \\
            \text{E}
            \end{array}$ & $\begin{array}{c}
            \text{O} \\
            \text{E} \\
            \text{E}
            \end{array}$ & $T_1$ \\[20pt] 
6 & $\left(\begin{array}{c}
            O_{4142}^- \\
            O_{2421}^- \\
            O_{1214}^-
            \end{array}\right)$ & $\begin{array}{c}
            \text{E} \\
            \text{E} \\
            \text{O}
            \end{array}$ & $\begin{array}{c}
            \text{E} \\
            \text{O} \\
            \text{E}
            \end{array}$ & $\begin{array}{c}
            \text{E} \\
            \text{E} \\
            \text{E}
            \end{array}$ & $\begin{array}{c}
            \text{O} \\
            \text{E} \\
            \text{E}
            \end{array}$ & $T_1$ \\[20pt] 
7 & $\left(\begin{array}{c}
            O_{3132}^+ \\
            O_{3431}^+ \\
            O_{3234}^+
            \end{array}\right)$ & $\begin{array}{c}
            \text{O} \\
            \text{O} \\
            \text{E}
            \end{array}$ & $\begin{array}{c}
            \text{O} \\
            \text{E} \\
            \text{O}
            \end{array}$ & $\begin{array}{c}
            \text{E} \\
            \text{E} \\
            \text{E}
            \end{array}$ & $\begin{array}{c}
            \text{E} \\
            \text{O} \\
            \text{O}
            \end{array}$ & $T_2$ \\[20pt] 
8 & $\left(\begin{array}{c}
            O_{4142}^+ \\
            O_{2421}^+ \\
            O_{1214}^+
            \end{array}\right)$ & $\begin{array}{c}
            \text{O} \\
            \text{O} \\
            \text{E}
            \end{array}$ & $\begin{array}{c}
            \text{O} \\
            \text{E} \\
            \text{O}
            \end{array}$ & $\begin{array}{c}
            \text{E} \\
            \text{E} \\
            \text{E}
            \end{array}$ & $\begin{array}{c}
            \text{E} \\
            \text{O} \\
            \text{O}
            \end{array}$ & $T_2$ \\[20pt] 
9 & $\left(\begin{array}{c}
            O_{3212}^+ + O_{3414}^+ \\
            O_{3121}^+ + O_{3424}^+ \\
            O_{3141}^+ + O_{3242}^+
            \end{array}\right)$ & $\begin{array}{c}
            \text{O} \\
            \text{E} \\
            \text{E}
            \end{array}$ & $\begin{array}{c}
            \text{E} \\
            \text{O} \\
            \text{E}
            \end{array}$ & $\begin{array}{c}
            \text{O} \\
            \text{O} \\
            \text{O}
            \end{array}$ & $\begin{array}{c}
            \text{E} \\
            \text{E} \\
            \text{O}
            \end{array}$ & $T_1$ \\[20pt] 
10 & $\left(\begin{array}{c}
            O_{3212}^- + O_{3414}^- \\
            O_{3121}^- + O_{3424}^- \\
            O_{3141}^- + O_{3242}^-
            \end{array}\right)$ & $\begin{array}{c}
            \text{E} \\
            \text{O} \\
            \text{O}
            \end{array}$ & $\begin{array}{c}
            \text{O} \\
            \text{E} \\
            \text{O}
            \end{array}$ & $\begin{array}{c}
            \text{O} \\
            \text{O} \\
            \text{O}
            \end{array}$ & $\begin{array}{c}
            \text{O} \\
            \text{O} \\
            \text{E}
            \end{array}$ & $T_1$ \\[20pt] 
11 & $\left(\begin{array}{c}
            O_{3212}^+ - O_{3414}^+ \\
            O_{3121}^+ - O_{3424}^+ \\
            O_{3141}^+ - O_{3242}^+
            \end{array}\right)$ & $\begin{array}{c}
            \text{O} \\
            \text{E} \\
            \text{E}
            \end{array}$ & $\begin{array}{c}
            \text{E} \\
            \text{O} \\
            \text{E}
            \end{array}$ & $\begin{array}{c}
            \text{O} \\
            \text{O} \\
            \text{O}
            \end{array}$ & $\begin{array}{c}
            \text{E} \\
            \text{E} \\
            \text{O}
            \end{array}$ & $T_2$ \\[20pt] 
12 & $\left(\begin{array}{c}
            O_{3212}^- - O_{3414}^- \\
            O_{3121}^- - O_{3424}^- \\
            O_{3141}^- - O_{3242}^-
            \end{array}\right)$ & $\begin{array}{c}
            \text{E} \\
            \text{O} \\
            \text{O}
            \end{array}$ & $\begin{array}{c}
            \text{O} \\
            \text{E} \\
            \text{O}
            \end{array}$ & $\begin{array}{c}
            \text{O} \\
            \text{O} \\
            \text{O}
            \end{array}$ & $\begin{array}{c}
            \text{O} \\
            \text{O} \\
            \text{E}
            \end{array}$ & $T_2$ \\[20pt] 
13 & $O_{3124}^+ +O_{3241}^+ +O_{3412}^+$ & $\text{O}$ & $\text{O}$ & $\text{O}$ & $\text{O}$ & $A_1$ \\[5pt]
14 & $O_{3124}^- +O_{3241}^- +O_{3412}^-$ & $\text{E}$ & $\text{E}$ & $\text{O}$ & $\text{E}$ & $A_1$ \\[5pt]
15 & $\left(\begin{array}{c}
            2 O_{3412}^+ -O_{3241}^+ -O_{3124}^+ \\
            O_{3124}^+ -O_{3241}^+
            \end{array}\right)$ & $\text{O}$ & $\text{O}$ & $\text{O}$ & $\text{O}$ & $E$ \\[15pt]
16 & $\left(\begin{array}{c}
            2 O_{3412}^- -O_{3241}^- -O_{3124}^- \\
            O_{3124}^- -O_{3241}^-
            \end{array}\right)$ & $\text{E}$ & $\text{E}$ & $\text{O}$ & $\text{E}$ & $E$ \\[15pt] 

\end{tabular}
\caption{Groups of operators exhibiting different parity transformations, along with the corresponding representation of the octahedral group.}
\label{tb:parity_rotation_symmetries}
\end{table} 

\section{Perturbative Calculation - Results}
\label{sec:results}

\begin{figure}[h]
    \centering
    \includegraphics[width=0.8\textwidth]{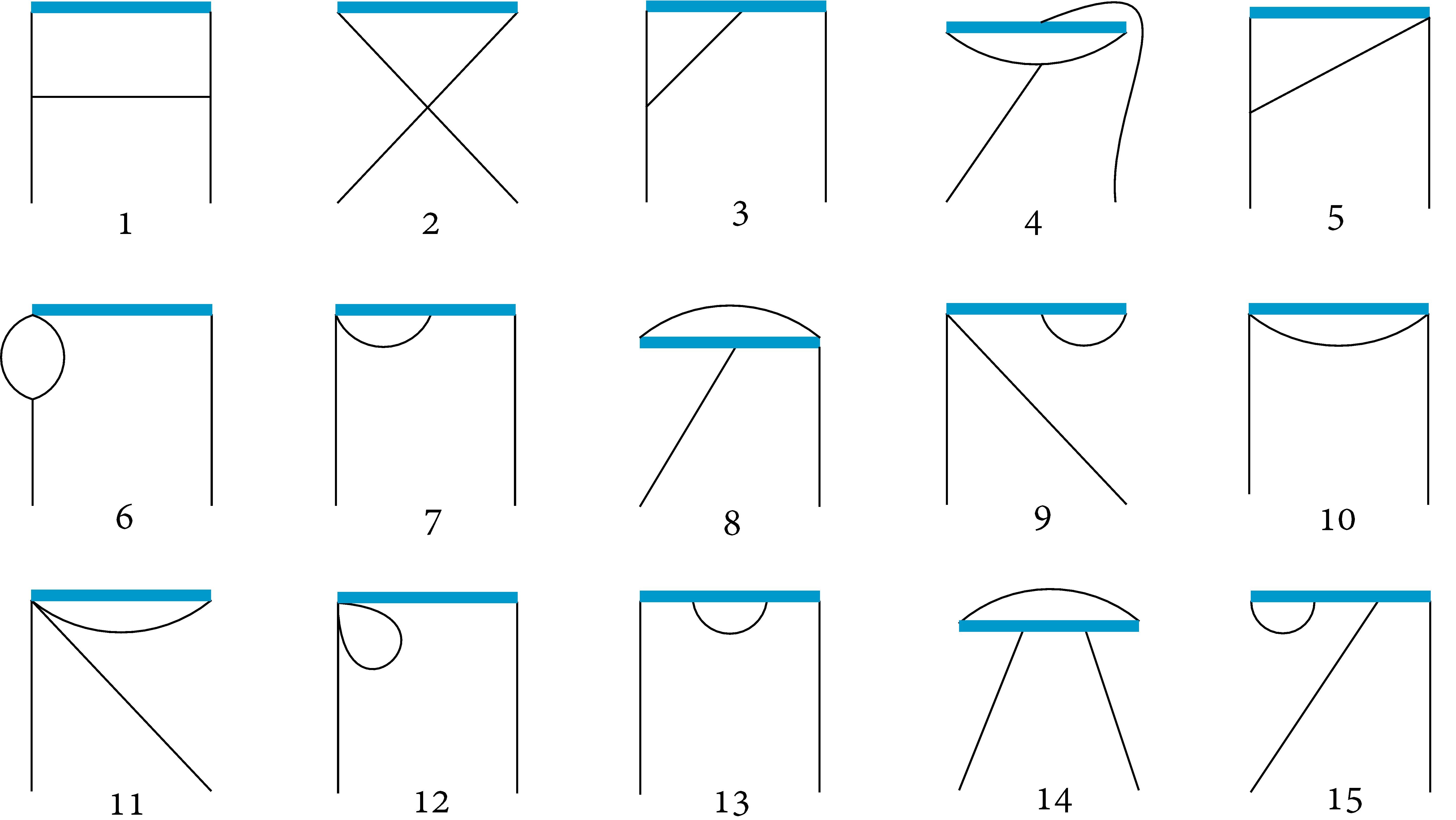}
    \caption{Feynman diagrams contributing to the one-loop calculation of the Green’s functions of the nonlocal operators. Mirror diagrams are not shown, for compactness. Solid lines represent gluons. The operator insertion is denoted by a solid box.}
    \label{fig:Feynman_diagrams}
\end{figure}

In this section, we provide the one-loop results for the $\overline{\rm MS}$ renormalized Green's functions of the operators, along with the renormalization factors in $\overline{\rm MS}$ and the conversion factors between the ${\rm RI}'$ and $\overline{\rm MS}$ schemes. Our calculations have been performed in both DR and LR. Due to the very lengthy expressions of the renormalized Green's functions and conversion factors, we include them as Supplemental Material, provided in the form of a \textit{Mathematica} input file named ``\texttt{Renormalized\_Greens\_Functions\_and\_Conversion\_Factors.m}". The Feynman diagrams corresponding to the one-loop two-point Green's functions $\Lambda_{O}$ of Eq.~\eqref{eq:Green_f_nonlocal} are illustrated in Fig.~\ref{fig:Feynman_diagrams}. Diagrams 1 and 2 contain two gluon fields stemming from the operator insertion (denoted as a blue-filled rectangle), while diagrams 3 to 6 (7 to 15) contain three (four) gluon fields stemming from the operator, which can be from either side (i.e., emerging from either $F_{\mu \nu}$) or from the center (i.e., originating from the Wilson line) of the operator.

Compared to Refs.~\cite{Wang:2019tgg,Wang:2017qyg} some additional diagrams are present in Fig.~\ref{fig:Feynman_diagrams}. In the Green's functions which we have calculated we have made no assumptions on the values of the Lorentz indices for the external gluons and for the operator; further, we have worked with a generic gauge and with off-shell gluons. A consequence of this is that certain diagrams (4, 8, 14) give non-vanishing contributions to these Green's functions. Another two diagrams (9, 15) have been included for completeness, but they contribute zero to both DR and LR.
Finally, diagrams 11 and 12 also contribute to LR, even though they vanish in DR. 

\subsection{Dimensional Regularization}

We first present our results from DR. The computations are performed in $D$-dimensional Euclidean spacetime, where $D=4-2\epsilon$ and $\epsilon$ is the regularization parameter. In contrast to two-point Green's functions involving local operators, the integration results become significantly more complicated due to the presence of both the external momentum $q$ and the length of the Wilson line $z$ in the integrands. Additionally, there is a nontrivial dependence on the preferred direction of the Wilson line, leading to further complexity. We apply new techniques,  similar to \cite{Spanoudes2018}, for handling one-loop tensor integrals with an exponential factor in $D$ dimensions. For the elimination of the poles in $\epsilon$, we adopt the $\overline{\rm MS}$ scheme.  

\subsubsection{Renormalization Functions}

We start by considering the amputated tree-level Green’s functions of Eq.~(\ref{eq:Green_f_nonlocal}), which read:
\begin{equation}
    \begin{aligned}
        \Lambda^{\text{tree}}_{O_{\mu \nu \rho \sigma}}= \delta^{a b} 
        \big( 
        &+ q_{ \mu} q_{ \rho} \ \delta_{\alpha \nu} \delta_{\beta \sigma} \ e^{-iz q_3}
        + q_{\mu} q_{\rho} \ \delta_{\alpha \sigma} 
        \delta_{\beta \nu} \ e^{iz q_3}\\
        &- q_{\nu} q_{\rho} \ \delta_{\alpha \mu} \delta_{\beta \sigma} \ e^{-iz q_3}
        - q_{\nu} q_{\rho} \ \delta_{\alpha \sigma} \delta_{\beta \mu} \ e^{iz q_3}\\
        &- q_{\mu} q_{\sigma} \ \delta_{\alpha \nu} \delta_{\beta \rho} \ e^{-iz q_3}
        - q_{\mu} q_{\sigma} \ \delta_{\alpha \rho} \delta_{\beta \nu} \ e^{iz q_3}\\
        &+ q_{\nu} q_{\sigma} \ \delta_{\alpha \mu} \delta_{\beta \rho} \ e^{-iz q_3}
        + q_{\nu} q_{\sigma} \ \delta_{\alpha \rho} \delta_{\beta \mu} \ e^{iz q_3}
        \big)
    \end{aligned}
    \label{eq:Tree_level_green_function}
\end{equation} 
where $z$ is the length of the Wilson line. Notice that, as expected, the above expression is antisymmetric in $\{\mu , \nu\}$ and $\{\rho , \sigma\}$; also, it is symmetric under $(\mu,\nu) \leftrightarrow (\rho,\sigma)$ and under $(\alpha,\beta,q) \leftrightarrow (\beta,\alpha,-q)$. 

Subsequently, we proceed to the one-loop calculations. In DR, diagrams 11, 12 do not exist. We ﬁnd that only diagrams 3, 6, and 13, contribute to the $1/\epsilon$ terms, and therefore the renormalization function of the operators in $\overline{\rm MS}$ is not affected by the remaining diagrams. However, they contribute to the renormalized Green’s functions and the conversion factors. Below we present the $\mathcal{O}(1/\epsilon)$ contributions of the perturbative calculation:
\begin{equation}
    \left. \Lambda^{\text{1-loop}}_{O_{\mu \nu \rho \sigma}}\right|_{\mathcal{O}(1/\epsilon)}= \frac{g_0^2 N_c}{16 \epsilon \pi^2}
    \left( \delta_{\mu 3}+\delta_{\nu 3}+\delta_{\rho 3}+\delta_{\sigma 3}- \frac{1}{2} \beta \right) \Lambda_O^{\text{tree}}
    \label{eq:oneloop_nonlocal}
\end{equation} 
The computation was carried out in an arbitrary covariant gauge, allowing for a direct verification of the gauge invariance of the renormalization factors. It should be noted that, at the one-loop level in DR, the pole terms are proportional to the tree-level values for each one of the operators, indicating no mixing with operators of equal dimension. 

The renormalization factor of the gluon field in (DR, $\overline{\rm MS}$) is given by~\cite{Gracey2003}:
\begin{equation}
    Z_A^{\text{DR}, \overline{\rm MS}}= 1+ \frac{g^2}{16 \epsilon \pi^2} 
    \left( \frac{13 N_c}{6} - \frac{N_c}{2}\left(1-\beta\right) - \frac{2}{3} N_f  \right)
    \label{eq:renorm_factor_gluon}
\end{equation} 
Using the $\overline{\rm MS}$ renormalization condition and Eqs.~(\ref{eq:Green_f_renormalized}), ~(\ref{eq:oneloop_nonlocal}), and ~(\ref{eq:renorm_factor_gluon}), the renormalization function of the operators turns out to be diagonal, both in the
original basis ($O_{\mu \nu \rho \sigma}$) and in the basis of Table~\ref{tb:parity_rotation_symmetries}. Its value is:
\begin{equation}
    Z^{\rm DR, \overline{MS}}_{O_{\mu \nu \rho \sigma}}= 1+ \frac{g^2}{16 \epsilon \pi^2} \left( \left( \frac{5}{3} + \delta_{\mu 3}+\delta_{\nu 3}+\delta_{\rho 3}+\delta_{\sigma 3}\right) N_c- \frac{2}{3} N_f  \right) 
    \label{eq:renorm_factor_operator}
\end{equation} 
[We recall that the cases $\mu=\nu$ and $\rho=\sigma$ give a vanishing operator; thus, it is understood that $\mu \neq \nu$ and $\rho \neq \sigma$ in Eq.~(\ref{eq:renorm_factor_operator}).] We observe that this result depends on the choice of indices for the operators, specifically whether they align with the direction of the Wilson line or not. In the basis of Table~\ref{tb:parity_rotation_symmetries}, the diagonal matrix $Z^{\rm DR, \rm \overline{MS}}_{ij}$ takes the form: 
\begin{equation}
    Z^{\text{DR}, \overline{\rm MS}}_{ij}= \delta_{ij} \left[ 1+ \frac{g^2}{16 \epsilon \pi^2} \left( \left( \frac{5}{3} + \omega_{i} \right) N_c- \frac{2}{3} N_f  \right) \right]
    \label{eq:renorm_factor_operator_new_basis}
\end{equation} 
where $\omega_{i}$ is defined as follows:
\begin{equation}
\omega_{i} = 
\begin{cases} 
0 & \text{for } i = 2,4,6,8 \\
1 & \text{for } i = 9 \text{-}16 \\
2 & \text{for } i = 1,3,5,7 
\end{cases}
\label{eq:omega}
\end{equation}
At this point, we remind the reader that groups containing multiplets share the same renormalization factor for each component within the multiplet. Note that Eq.~\ref{eq:renorm_factor_operator_new_basis} is compatible with rotational symmetry arguments in the continuum described by Eq.~\ref{eq:rotations_continuum_Z}. Our results agree with previous studies using the auxiliary-field formulation~\cite{Dorn:1981wa,Zhang2018,Braun:2020ymy}.

As expected from gauge invariance in $\overline{\rm MS}$, the $\beta$ dependence disappears in the renormalization function of the operators, upon taking into account the gluon field renormalization function. Gauge invariance cannot be ensured in all schemes due to the presence of gauge-dependent renormalized external fields in the Green’s functions.

It is worth mentioning that the renormalization function of the operators is independent of the length of the Wilson line ($z$). There is no dimensionless factor dependent on $z$ that could emerge in the pole part because the leading pole at each loop cannot depend on external momenta or the renormalization scale. Consequently, $z$ independence is expected to persist at all orders in perturbation theory.

\subsubsection{\texorpdfstring{${\rm RI}'$}{TEXT} renormalization prescription}
\label{subsec:RIprime}

In a ${\rm RI}'$ scheme, there exists significant flexibility in defining normalization conditions in Green’s functions, particularly in cases involving operator mixing. These variations only differ by finite terms. Hence, it is convenient to adopt a minimal prescription, containing the smallest possible set of operators prone to mixing, typically consistent with the mixing pattern identified by symmetries. This includes groups $\{1,\ 2\}$, $\{3,\ 4\}$, $\{5,\ 6\}$, $\{7,\ 8\}$ of Table~\ref{tb:parity_rotation_symmetries}. However, such a scheme must be independent of the regularization method, incorporating any potential additional finite or power-divergent mixing, as encountered, for instance, in lattice regularization.

A practical choice for a ${\rm RI}'$-like scheme suitable for nonperturbative studies is to consider four $2\times2$ mixing matrices, since there are four mixing pairs of operators, alongside eight $1\times1$ matrices for operators that are multiplicatively renormalizable. However, renormalization conditions for operator 13 cannot be set, as the bare Green's function under study is zero. To properly select its renormalization conditions, further calculations involving other Green's functions, such as three-point Green's functions, are required. Consequently, we need to impose a total of 23 conditions to identify the elements of these matrices. The proposed renormalization conditions for this variant of the ${\rm RI}'$ scheme are as follows (where $\alpha$ and $\beta$ are the Lorentz indices of the external gluons, cf. Eq.~\ref{eq:Green_f_nonlocal}):
\begin{equation}
     \frac{{\rm Tr} \left[ \Lambda_{O_{(i)}}^{\mathrm{RI}^{\prime}}(\bar{q},z) \right]}{N_c^2-1} \Bigg\rvert_{\substack{\bar{q}_3=\bar{q}_4=0, \\ \alpha=\beta=3}} = \frac{{\rm Tr} \left[ \Lambda_{O_{(i)}}^{\mathrm{tree}}(\bar{q},z) \right]}{N_c^2-1}  \Bigg\rvert_{\substack{\bar{q}_3=\bar{q}_4=0, \\ \alpha=\beta=3}}  =  
    \begin{cases} 
    2 \bar{q}^2 & \text{for } i = 1 \\
    - 2 \bar{q}^2 & \text{for } i = 3 \\
    2 \bar{q}_1 \bar{q}_2 & \text{for } i = 7 \\
    0 & \text{for } i = 2,4,8 \\
    \end{cases}\hspace{30pt}
    \label{eq:RIprime_1}
\end{equation}

\begin{equation}
     \frac{{\rm Tr} \left[ \Lambda_{O_{(i)}}^{\mathrm{RI}^{\prime}}(\bar{q},z) \right]}{N_c^2-1} \Bigg\rvert_{\substack{\bar{q}_3=\bar{q}_4=0, \\ \alpha=\beta=4}} = \frac{{\rm Tr} \left[ \Lambda_{O_{(i)}}^{\mathrm{tree}}(\bar{q},z) \right]}{N_c^2-1}  \Bigg\rvert_{\substack{\bar{q}_3=\bar{q}_4=0, \\ \alpha=\beta=4}}  =  
    \begin{cases} 
    2 \bar{q}^2 & \text{for } i = 2 \\
    - 2 \bar{q}^2 & \text{for } i = 4 \\
    2 \bar{q}_1 \bar{q}_2 & \text{for } i = 8 \\
    0 & \text{for } i = 1,3,7 \\
    \end{cases}\hspace{30pt}
    \label{eq:RIprime_2}
\end{equation}

\begin{equation}
     \frac{{\rm Tr} \left[ \Lambda_{O_{(i)}}^{\mathrm{RI}^{\prime}}(\bar{q},z) \right]}{N_c^2-1} \Bigg\rvert_{\substack{\bar{q}_1=0, \\\alpha=1, \beta=3}} = \frac{{\rm Tr} \left[ \Lambda_{O_{(i)}}^{\mathrm{tree}}(\bar{q},z) \right]}{N_c^2-1}  \Bigg\rvert_{\substack{\bar{q}_1=0, \\\alpha=1, \beta=3}}  =  
    \begin{cases} 
    i \sin{(z \bar{q}_3)}\, \bar{q}_2 \bar{q}_3 & \text{for } i = 5 \\
    0 & \text{for } i = 6 
    \end{cases} \hspace{10pt}
    \label{eq:RIprime_3}
\end{equation}

\begin{equation}
     \frac{{\rm Tr} \left[ \Lambda_{O_{(i)}}^{\mathrm{RI}^{\prime}}(\bar{q},z) \right]}{N_c^2-1} \Bigg\rvert_{\substack{\bar{q}_1=0, \\ \alpha=1, \beta=4}} = \frac{{\rm Tr} \left[ \Lambda_{O_{(i)}}^{\mathrm{tree}}(\bar{q},z) \right]}{N_c^2-1}  \Bigg\rvert_{\substack{\bar{q}_1=0, \\ \alpha=1, \beta=4}}  = 
    \begin{cases} 
    0 & \text{for } i = 5 \\
    i \sin{(z \bar{q}_3)}\, \bar{q}_2 \bar{q}_4 & \text{for } i = 6 
    \end{cases} \hspace{10pt}
    \label{eq:RIprime_4}
\end{equation}

\begin{equation}
     \frac{{\rm Tr} \left[ \Lambda_{O_{(i)}}^{\mathrm{RI}^{\prime}}(\bar{q},z) \right]}{N_c^2-1} \Bigg\rvert_{\substack{\bar{q}_3=\bar{q}_4=0, \\ \alpha=1,\beta=3}} = \frac{{\rm Tr} \left[ \Lambda_{O_{(i)}}^{\mathrm{tree}}(\bar{q},z) \right]}{N_c^2-1}  \Bigg\rvert_{\substack{\bar{q}_3=\bar{q}_4=0, \\ \alpha=1,\beta=3}}  =  \bar{q}_2^2 \quad\quad \text{for } i = 9,11 \hspace{48pt}
    \label{eq:RIprime_5}
\end{equation}

\begin{equation}
     \frac{{\rm Tr} \left[ \Lambda_{O_{(i)}}^{\mathrm{RI}^{\prime}}(\bar{q},z) \right]}{N_c^2-1} \Bigg\rvert_{\substack{\bar{q}_1=\bar{q}_4=0, \\ \alpha=1,\beta=3}} = \frac{{\rm Tr} \left[ \Lambda_{O_{(i)}}^{\mathrm{tree}}(\bar{q},z) \right]}{N_c^2-1}  \Bigg\rvert_{\substack{\bar{q}_1=\bar{q}_4=0, \\ \alpha=1,\beta=3}}  =  i \sin{(z \bar{q}_3)}\, \bar{q}_2^2 \quad\quad \text{for } i = 10,12 
    \label{eq:RIprime_6}
\end{equation}

\begin{equation}
     \frac{{\rm Tr} \left[ \Lambda_{O_{(i)}}^{\mathrm{RI}^{\prime}}(\bar{q},z) \right]}{N_c^2-1} \Bigg\rvert_{\substack{\bar{q}_1=\bar{q}_4=0, \\ \alpha=4,\beta=1}} = \frac{{\rm Tr} \left[ \Lambda_{O_{(i)}}^{\mathrm{tree}}(\bar{q},z) \right]}{N_c^2-1}  \Bigg\rvert_{\substack{\bar{q}_1=\bar{q}_4=0, \\ \alpha=4,\beta=1}}  =  
     \begin{cases} 
    2i \sin{(z \bar{q}_3)}\, \bar{q}_2 \bar{q}_3 & \text{for } i = 14 \\
    i \sin{(z \bar{q}_3)}\, \bar{q}_2 \bar{q}_3 & \text{for } i = 16 
    \end{cases} 
    \label{eq:RIprime_7}
\end{equation}

\begin{equation}
     \frac{{\rm Tr} \left[ \Lambda_{O_{(i)}}^{\mathrm{RI}^{\prime}}(\bar{q},z) \right]}{N_c^2-1} \Bigg\rvert_{\substack{\bar{q}_3=\bar{q}_4=0, \\ \alpha=3,\beta=4}} = \frac{{\rm Tr} \left[ \Lambda_{O_{(i)}}^{\mathrm{tree}}(\bar{q},z) \right]}{N_c^2-1}  \Bigg\rvert_{\substack{\bar{q}_3=\bar{q}_4=0, \\ \alpha=3,\beta=4}}  = \frac{\bar{q}_1 \bar{q}_2}{2} \quad\quad \text{for } i = 15 \hspace{45pt}
    \label{eq:RIprime_8}
\end{equation}
where the momentum of the external gluon fields is represented by $q_\nu$, while the four-vector $\bar{q}_\nu$ denotes the ${\rm RI}'$ renormalization scale. The trace in the above equations is performed across color space. It is important to note that considering only the magnitude of $\bar{q}$ doesn't fully define the renormalization prescription. Various directions within $\bar{q}$ correspond to distinct renormalization schemes, interconnected through finite renormalization factors. In our proposed conditions, we select certain values for the Lorentz indices $\alpha$, $\beta$ and we set specific components of $\bar{q}$ to zero. With this choice, we isolate, in each condition, one of the possible Lorentz structures appearing in the Green's functions $\Lambda_{O(i)}$, in a way as to lead to a solvable system of conditions and to, as much as possible, simpler expressions. Other options can be tested by using our results for the Green's functions provided in the supplemental file.

For `minus-type' operators (i.e., for mixing pair $\{5,6\}$ and operators $10$,$12$,$14$, and $16$ with multiplicative renormalization) we cannot select $\bar{q}_3=0$ (nor $\bar{q}_3=\frac{\pi}{z} n$, where $n$ is an integer) because $\sin(\bar{q}_3 z)$ which appears in their tree-level expression will vanish, thus making this expression noninvertible. 

Note that the RI$'$ conditions are expressed in terms of amputated Green's functions. Consequently, in order to treat nonperturbative, non-amputated Green’s functions, coming from lattice simulations, we must multiply each external gluon by an inverse gluon propagator. Such a propagator is non-invertible in the Landau gauge, which is typically employed in simulations, however its inverse in the transverse subspace can be constructed in standard fashion, using singular value decomposition.

\subsubsection{Conversion factors}

The $\overline{\rm MS}$-renormalized Green's functions are the fundamental ingredient for the construction of the conversion factors between $\overline{\rm MS}$ and $\rm RI'$ scheme, defined in Eqs.~(\ref{eq:RIprime_1})--(\ref{eq:RIprime_8}). These renormalized Green's functions are equal to the finite part of $\Lambda_O^{\text{1-loop}}$ and are complex expressions involving integrals over Bessel functions. By applying the renormalization conditions of the $\rm RI'$ scheme in Eq.~(\ref{eq:Greens_function_MSbar_to_RIprime}), one can straightforwardly derive the $2\times2$ conversion factors for the mixing pairs of operators found in Section~\ref{sec:symmetries}, represented as $\mathcal{C}_{\{i,j\}}^{\rm \overline{MS}, RI'}$, where $i$ and $j$ denote the two operators belonging to a mixing pair. Due to the very lengthy expressions of the conversion factors, we present below only the explicit results for `plus-type' operators (i.e., for mixing pairs $\{1,2\}$,$\{3,4\}$,$\{7,8\}$ and multiplicatively renormalizable operators $9$, $11$, and $15$); the expressions are presented for a general gauge-fixing parameter ($\beta$) in terms of the quantities $F_{1}(\bar{q}^2,\bar{q}_3, z)$--$F_{9}(\bar{q}^2,\bar{q}_3, z)$ where $\bar{q}$ is the 4-vector renormalization scale dictated by the renormalization conditions of RI$'$ for each operator set. The quantities $F_i$ are integrals of modified Bessel functions of the second kind, $K_{n}$, over a Feynman parameter, and are provided in Eqs.~(\ref{eq:BesselK_integrals_F1})--(\ref{eq:BesselK_integrals_F9}) of Appendix~\ref{ap:Feynman_parameter_integrals}. 

The expressions for the `minus-type' operators are provided in the supplementary \textit{Mathematica} input file. They involve a double integral of modified Bessel functions, see Appendix~\ref{ap:Feynman_parameter_integrals} for an example. 

Also, it is important to note that the conversion factors depend on the dimensionless quantities $z \overline{q}$ and $\overline{q}/\overline{\mu}$. The $\rm RI'$ and $\overline{\rm MS}$ renormalization scales ($\overline{q}$ and $\overline{\mu}$, respectively) have been left arbitrary. [The $\MSbar$ renormalization scale $\bar{\mu}$ stems from the renormalization of the coupling constant in D dimensions: $g = \mu^{-\epsilon} Z_g^{-1} g_0$, where $g_0$ ($g$) is the bare (renormalized) coupling constant and $\mu = \bar \mu \sqrt{e^{\gamma_E}/ 4\pi}$).]

\begin{align}
\begin{split}
    \left[\mathcal{C}_{\{1,2\}}^{\rm \overline{MS}, RI'}\right]_{1,1} =
    1 + \frac{g^2 N_c}{16 \pi^2} \Bigg[ 
    & \frac{67}{9} -\frac{\beta ^2}{4} + (\beta +2) \left(2 \gamma_E - \log (4) + \log \left(z^2 \bar{q}^2\right) \right) +\frac{\beta -4}{2} \log \left(\frac{\bar{q}^2}{\bar{\mu }^2}\right)  \\
    & +\frac{F_2-F_3}{2} \left(261 \beta+40\right) +(2-69 \beta ) \left(F_8-F_9\right)+\frac{F_7}{2} (23 \beta -1)  + \frac{F_1}{2} \left(-28 \beta - 11\right)  \\
    & + \bar{q}^2 | z| ^2 \frac{7 \beta}{2}  \left(3  \left(F_3-2 F_4+F_5\right)- \left(F_2-F_3\right)\right)+ \frac{6}{\bar{q}^2 | z| ^2} \left(1 - \beta + F_7 \right)  -\frac{10 N_f}{9 N_c} \Bigg]
    \label{eq:conversion_factor_12_11}
\end{split}
\end{align}

\begin{align}
\begin{split}
    \left[\mathcal{C}_{\{1,2\}}^{\rm \overline{MS}, RI'}\right]_{1,2} =
    \frac{g^2 N_c}{16 \pi^2} \Bigg[ 
    \frac{12}{\bar{q}^2 | z| ^2}\left( (2 \beta -1) F_7+1 \right) +4 (\beta -2) \left(F_2-F_3\right)+4(5 \beta -1) \left(F_8-F_9\right)+4 F_1 \Bigg]
    \label{eq:conversion_factor_12_12}
\end{split}
\end{align}

\begin{align}
\begin{split}
    \left[\mathcal{C}_{\{1,2\}}^{\rm \overline{MS}, RI'}\right]_{2,1} =
    \frac{g^2 N_c}{16 \pi^2} \Bigg[ 
    & \frac{37 \beta}{2} \left(F_2-F_3\right)+ \frac{F_7}{2} \left(-\frac{\beta }{2}-5\right) +(-5 \beta -2) \left(F_8-F_9\right)+\frac{5 F_1}{2} \\
    & + \bar{q}^2 | z| ^2 \frac{\beta}{2} \left(\frac{1}{2} \left(F_2-F_3\right)+ 5 \left(F_3-2 F_4+F_5\right)\right)+\frac{2}{\bar{q}^2 | z| ^2} \left( (\beta +3)- F_7 \right)\Bigg]
    \label{eq:conversion_factor_12_21}
\end{split}
\end{align}

\begin{align}
\begin{split}
    \left[\mathcal{C}_{\{1,2\}}^{\rm \overline{MS}, RI'}\right]_{2,2} =
    1 + \frac{g^2 N_c}{16 \pi^2} \Bigg[
    & \frac{31}{9}-\frac{\beta ^2}{4} +(\beta +2) \left( 2 \gamma_E - \log (4) + \log \left(z^2 \bar{q}^2\right) \right) +\frac{\beta}{2}   \log \left(\frac{\bar{q}^2}{\bar{\mu }^2}\right) \\
    & +(76 \beta +14) \left(F_2-F_3\right)+\left(-105 \beta -4\right) \frac{F_8-F_9}{2}+\frac{3 F_7}{4} (13 \beta +2) -14 \beta  F_1 \\
    & + \bar{q}^2 | z| ^2 \frac{\beta}{2}  \left(13 \left(F_3-2 F_4+F_5\right)-5 \left(F_2-F_3\right)\right)+\frac{4}{\bar{q}^2 | z| ^2} \left(4- \beta -14 \beta  F_7\right) -\frac{10 N_f}{9 N_c}\Bigg]
    \label{eq:conversion_factor_12_22}
\end{split}
\end{align}

\begin{align}
\begin{split}
    \left[\mathcal{C}_{\{3,4\}}^{\rm \overline{MS}, RI'}\right]_{1,1} =
    1 + \frac{g^2 N_c}{16 \pi^2} \Bigg[ 
    & \frac{67}{9} -\frac{\beta ^2}{4} + (\beta +2) \left(2 \gamma_E - \log (4) + \log \left(z^2 \bar{q}^2\right) \right) +\frac{\beta -4}{2} \log \left(\frac{\bar{q}^2}{\bar{\mu }^2}\right) \\
    & -7 (2 \beta +1) F_1+(69 \beta +8) \left(F_2-F_3\right) +\frac{F_7}{4} (49 \beta -2) +\left(-99 \beta -8\right) \frac{F_8-F_9}{2} \\
    & + \bar{q}^2 | z| ^2 \frac{7 \beta}{2}  \left(3  \left(F_3-2 F_4+F_5\right)- \left(F_2-F_3\right)\right) -\frac{10 N_f}{9 N_c} \Bigg]
    \label{eq:conversion_factor_34_11}
\end{split}
\end{align}

\begin{align}
\begin{split}
    \left[\mathcal{C}_{\{3,4\}}^{\rm \overline{MS}, RI'}\right]_{1,2} =
    \frac{g^2 N_c}{16 \pi^2} \Bigg[ 2(5 \beta -1) \left(F_2-F_3\right)+2(1-5 \beta ) \left(F_8-F_9\right)-2 F_1 \Bigg]
    \label{eq:conversion_factor_34_12}
\end{split}
\end{align}

\begin{align}
\begin{split}
    \left[\mathcal{C}_{\{3,4\}}^{\rm \overline{MS}, RI'}\right]_{2,1} =
    \frac{g^2 N_c}{16 \pi^2} \Bigg[ 
    & (-43 \beta -6) \left(F_2-F_3\right)+\frac{F_7}{2} (\beta -2) + 4(7 \beta +1) \left(F_8-F_9\right)+F_1 \\
    & +\bar{q}^2 | z| ^2 \frac{\beta}{2} \left(-  \left(F_2-F_3\right)-10  \left(F_3-2 F_4+F_5\right)\right) +\frac{4}{\bar{q}^2 | z| ^2}\left(F_7-\beta-3 \right) \Bigg]
    \label{eq:conversion_factor_34_21}
\end{split}
\end{align}

\begin{align}
\begin{split}
    \left[\mathcal{C}_{\{3,4\}}^{\rm \overline{MS}, RI'}\right]_{2,2} =
    1 + \frac{g^2 N_c}{16 \pi^2} \Bigg[ 
    & \frac{31}{9}-\frac{\beta ^2}{4} +(\beta +2) \left( 2 \gamma_E - \log (4) + \log \left(z^2 \bar{q}^2\right) \right) +\frac{\beta}{2}   \log \left(\frac{\bar{q}^2}{\bar{\mu }^2}\right) \\
    & -2 (7 \beta +3) F_1+(106 \beta +8) \left(F_2-F_3\right)+\frac{3 F_7}{4} (13 \beta +2) +\left(-105 \beta -4\right) \frac{F_8-F_9}{2} \\
    & + \bar{q}^2 | z| ^2 \frac{\beta}{2}  \left(13 \left(F_3-2 F_4+F_5\right)-5 \left(F_2-F_3\right)\right) +\frac{4}{\bar{q}^2 | z| ^2} \left(1 - \beta - (8 \beta +3) F_7 \right)-\frac{10 N_f}{9 N_c} \Bigg]
    \label{eq:conversion_factor_34_22}
\end{split}
\end{align}





\begin{align}
\begin{split}
    \left[\mathcal{C}_{\{7,8\}}^{\rm \overline{MS}, RI'}\right]_{1,1} =
    1 + \frac{g^2 N_c}{16 \pi^2} \Bigg[ 
    & \frac{67}{9} -\frac{\beta ^2}{4} + (\beta +2) \left(2 \gamma_E - \log (4) + \log \left(z^2 \bar{q}^2\right) \right) +\frac{\beta -4}{2} \log \left(\frac{\bar{q}^2}{\bar{\mu }^2}\right) \\
    &-7 (2 \beta +1) F_1+(69 \beta +8) \left(F_2-F_3\right) +\frac{F_7}{4} (49 \beta -2) +\left(-99 \beta -8\right) \frac{F_8-F_9}{2}  \\
    & + \bar{q}^2 | z| ^2 \frac{7 \beta}{2}  \left(3  \left(F_3-2 F_4+F_5\right)- \left(F_2-F_3\right)\right) -\frac{10 N_f}{9 N_c} \Bigg]
    \label{eq:conversion_factor_78_11}
\end{split}
\end{align}

\begin{align}
\begin{split}
    \left[\mathcal{C}_{\{7,8\}}^{\rm \overline{MS}, RI'}\right]_{1,2} =
    \frac{g^2 N_c}{16 \pi^2} \Bigg[ 2(1-5 \beta ) \left(F_2-F_3\right)+2(5 \beta -1) \left(F_8-F_9\right)+2 F_1 \Bigg]
    \label{eq:conversion_factor_78_12}
\end{split}
\end{align}

\begin{align}
\begin{split}
    \left[\mathcal{C}_{\{7,8\}}^{\rm \overline{MS}, RI'}\right]_{2,1} =
    \frac{g^2 N_c}{16 \pi^2} \Bigg[ 2(1-5 \beta ) \left(F_2-F_3\right)+6 \beta  \left(F_8-F_9\right)+2 F_1-2 F_7 \Bigg]
    \label{eq:conversion_factor_78_21}
\end{split}
\end{align}

\begin{align}
\begin{split}
    \left[\mathcal{C}_{\{7,8\}}^{\rm \overline{MS}, RI'}\right]_{2,2} =
    1 + \frac{g^2 N_c}{16 \pi^2} \Bigg[ 
    & \frac{31}{9}-\frac{\beta ^2}{4} +(\beta +2) \left( 2 \gamma_E - \log (4) + \log \left(z^2 \bar{q}^2\right) \right) +\frac{\beta}{2}   \log \left(\frac{\bar{q}^2}{\bar{\mu }^2}\right) \\ 
    & +(-14 \beta -3) F_1+(69 \beta +8) \left(F_2-F_3\right) +\frac{F_7}{4} (41 \beta +6) +\left(-95 \beta-4\right) \frac{F_8-F_9}{2} \\ 
    & + \bar{q}^2 | z| ^2 \frac{\beta}{2}  \left(13 \left(F_3-2 F_4+F_5\right)-5 \left(F_2-F_3\right)\right) -\frac{10 N_f}{9 N_c} \Bigg]
    \label{eq:conversion_factor_78_22}
\end{split}
\end{align}

\begin{align}
\begin{split}
    \mathcal{C}_{\{9\}}^{\rm \overline{MS}, RI'} =
    1 + \frac{g^2 N_c}{16 \pi^2} \Bigg[ 
    & \frac{49}{9}-\frac{\beta ^2}{4} + (\beta +2) \left(2 \gamma_E - \log (4) + \log \left(z^2 \bar{q}^2\right) \right) +\frac{\beta -2}{2} \log \left(\frac{\bar{q}^2}{\bar{\mu }^2}\right) \\
    & - 2 (8 \beta +1) F_1+(125 \beta -2) \left(F_2-F_3\right)+\frac{3 F_7}{4} (15 \beta +2) +\left(-153 \beta -4\right) \frac{F_8-F_9}{2} \\
    & + \bar{q}^2 | z| ^2 \frac{\beta}{2} \left(23   \left(F_3-2 F_4+F_5\right)-6 \left(F_2-F_3\right)\right)+\frac{4}{\bar{q}_2^2 | z| ^2} \left(1- \beta -3 F_7\right)\\ 
    & + \frac{\bar{q}^2}{\bar{q}_2^2} \left(2 (\beta -2) F_1+(12-13 \beta ) \left(F_2-F_3\right)  + 21 \beta  \left(F_8-F_9\right)-\frac{\beta  F_7}{2}\right)  -\frac{10 N_f}{9 N_c} \Bigg]
    \label{eq:conversion_factor_9}
\end{split}
\end{align}


\begin{align}
\begin{split}
    \mathcal{C}_{\{11\}}^{\rm \overline{MS}, RI'} =
    1 + \frac{g^2 N_c}{16 \pi^2} \Bigg[
    & \frac{49}{9}-\frac{\beta ^2}{4} + (\beta +2) \left(2 \gamma_E - \log (4) + \log \left(z^2 \bar{q}^2\right) \right) +\frac{\beta -2}{2} \log \left(\frac{\bar{q}^2}{\bar{\mu }^2}\right) \\
    & -7 (2 \beta +1) F_1+(79 \beta +6) \left(F_2-F_3\right) +\frac{3 F_7}{4} (15 \beta +2) +\left(-101 \beta-4\right) \frac{F_8-F_9}{2} \\
    & + \bar{q}^2 | z| ^2 \frac{\beta}{2} \left(23   \left(F_3-2 F_4+F_5\right)-6 \left(F_2-F_3\right)\right)-\frac{10 N_f}{9 N_c} \Bigg]
    \label{eq:conversion_factor_11}
\end{split}
\end{align}



\begin{align}
\begin{split}
    \mathcal{C}_{\{15\}}^{\rm \overline{MS}, RI'} =
    1 + \frac{g^2 N_c}{16 \pi^2} \Bigg[ 
    & \frac{49}{9}-\frac{\beta ^2}{4} + (\beta +2) \left(2 \gamma_E - \log (4) + \log \left(z^2 \bar{q}^2\right) \right) +\frac{\beta -2}{2} \log \left(\frac{\bar{q}^2}{\bar{\mu }^2}\right) \\
    & +(7 \beta +6) \left(F_2-F_3\right)+\left(45 \beta+6\right) \frac{F_7-F_8}{2}+\left(-17 \beta-4\right) \frac{F_8-F_9}{2}-4 F_1 \\
    & +144 \beta  \left(F_2-2 F_3+F_4\right)-84 \beta  \left(F_8-2 F_9+F_{10}\right)+(-28 \beta -6) \left(F_1-F_2\right) \\
    & +\bar{q}^2 | z| ^2 \frac{\beta}{2} \left(-12 \left(F_2-2 F_3+F_4\right)+ 35 \left(F_3-2 F_4+F_5\right)-24 \left(F_4-2 F_5+F_6\right)\right) -\frac{10 N_f}{9 N_c} \Bigg]
    \label{eq:conversion_factor_15}
\end{split}
\end{align}


Plotting conversion factors for the parameters used in lattice simulations can offer very useful insights and visual representations. To facilitate this, we select specific values of the free parameters that correspond to the $N_f = 2 + 1 + 1$ ensemble of twisted-mass clover-improved fermions described in Ref.~\cite{Delmar:2023agv}. In this setup, the $\overline{\rm MS}$ scale is fixed at $\overline{\mu} = 2 \ \text{GeV}$ while the lattice volume is $L^3 \times T$ with $L = 32$ and $T = 64$ (in lattice units). The lattice spacing is $a = 0.0938$ fm, while $g^2 = 3.47625$ and $\beta = 1$ (Landau gauge). 

The ${\rm RI}'$ scale is defined in lattice units as $a\bar{q} = \left( \frac{2\pi}{L} n_1, \frac{2\pi}{L} n_2, \frac{2\pi}{L} n_3, \frac{2\pi}{T} \left( n_4 + \frac{1}{2} \right) \right)$, where $n_i$ are integers. For the momentum scales, we choose isotropic spatial directions ($n_1 = n_2 = n_3$) when possible and introduce a nonzero twist of $1/2$ in the temporal component. This choice aligns with the antiperiodic boundary conditions applied to the fermion fields in the temporal direction. Additionally, we rescale the length of the Wilson line with the lattice spacing, denoted as $z/a$.

Depending on the choice of $\bar{q}$ the numerical values of the conversion factors can be excessively large. It is thus important to tune the values of $\bar{q}$ accordingly. Similarly to the continuum, most appropriate choices of values for $\bar{q}_3$ on the lattice will be: $\bar{q}_3=\frac{2 \pi}{a L} n_3$, where $n_3$ is an odd integer: these choices guarantee that tree-level Green's functions will be invertible for all integer values of $z/a$ in the range $1\leq z/a < L/2$.

\begin{figure}[h]
    \centering
    \begin{subfigure}{0.45\linewidth}
        \centering
        \includegraphics[width=\linewidth]{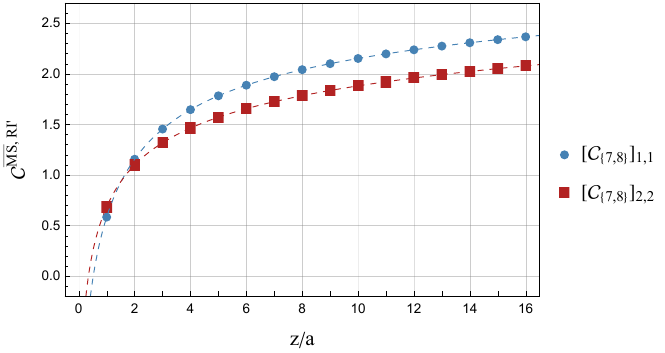}
        \caption{Diagonal elements.}
    \end{subfigure}
    \hfill
    \begin{subfigure}{0.45\linewidth}
        \centering
        \includegraphics[width=\linewidth]{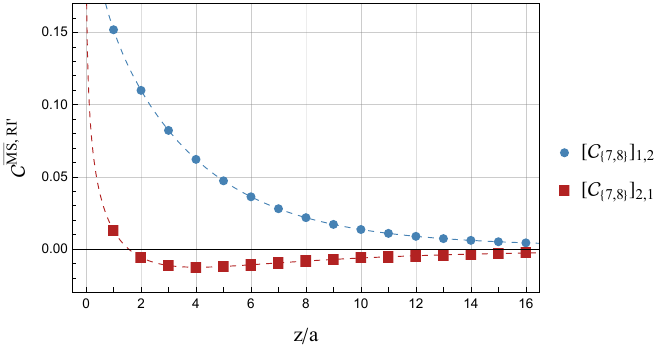}
        \caption{Nondiagonal elements.}
    \end{subfigure}
    \caption{Elements of $\mathcal{C}_{\{7,8\}}^{\rm \overline{MS}, {\rm RI}'}$ conversion matrix as a function of $z/a$.}
    \label{fig:conversion_factors_7_8}
\end{figure}

\begin{figure}[h]
    \centering
    \includegraphics[width=0.45\linewidth]{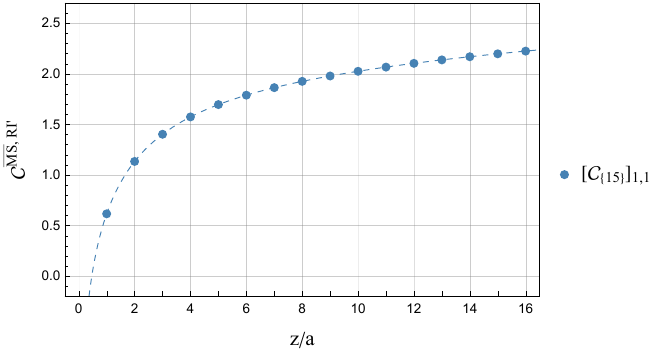}
    \caption{Conversion factor $\mathcal{C}_{\{15\}}^{\rm \overline{MS}, {\rm RI}'}$ as a function of $z/a$}
    \label{fig:conversion_factor_15}
\end{figure}

As an example, we apply the following values for the plus-type operators: $n_1 = n_2 = 3$, $n_3 = 0$, and $n_4 = -1/2$. The conversion matrix elements for pair $\{7,8\}$ are shown in Fig.~\ref{fig:conversion_factors_7_8}; the other plus-type mixing pairs, i.e., $\{1,2\}$ and $\{3,4\}$, have similar qualitative behavior. The plus-type operators undergoing multiplicative renormalization (9,11, and 15) exhibit a similar graphical representation as demonstrated, for example, in Fig. \ref{fig:conversion_factor_15}, for operator 15.

\begin{figure}[htbp]
    \centering
    \begin{subfigure}{0.45\linewidth}
        \centering
        \includegraphics[width=\linewidth]{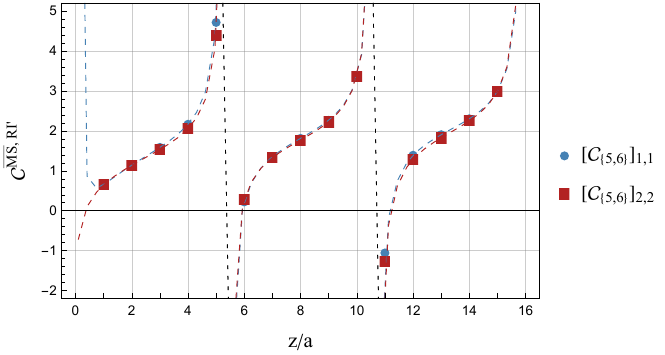}
        \caption{Diagonal elements.}
    \end{subfigure}
    \hfill
    \begin{subfigure}{0.45\linewidth}
        \centering
        \includegraphics[width=\linewidth]{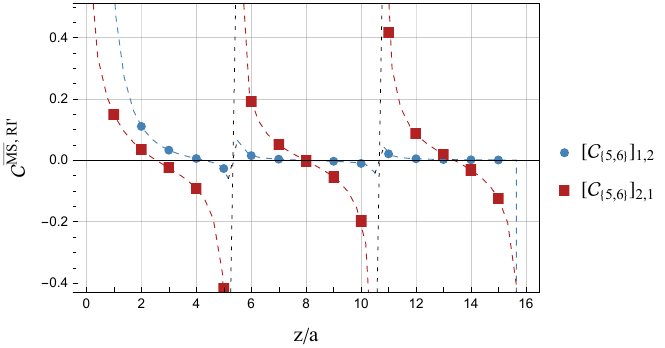}
        \caption{Nondiagonal elements.}
    \end{subfigure}
    \caption{Elements of $\mathcal{C}_{\{5,6\}}^{\rm \overline{MS}, {\rm RI}'}$ conversion matrix as a function of $z/a$.}
    \label{fig:conversion_factors_5_6}
\end{figure}

\begin{figure}
    \centering
    \includegraphics[width=0.45\linewidth]{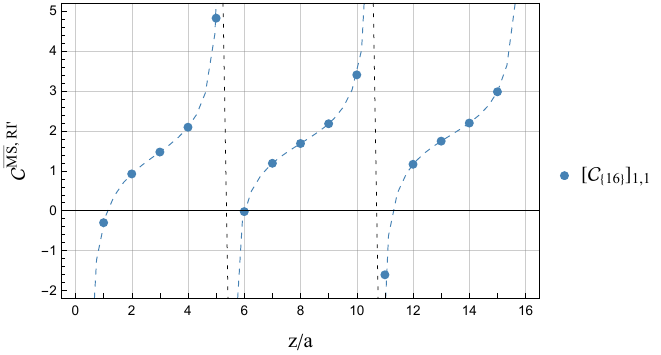}
    \caption{Conversion factor $\mathcal{C}_{\{16\}}^{\rm \overline{MS}, {\rm RI}'}$ as a function of $z/a$}
    \label{fig:conversion_factor_16}
\end{figure}

Furthermore, in Fig.\ref{fig:conversion_factors_5_6}, the conversion matrix elements for the pair $\{5,6\}$ of minus-type operators are presented, where the values $n_2 = n_3 = 3$, $n_1 = 0$, and $n_4 = 5$ are employed. For the remaining minus-type operators, we have selected $n_1 = n_2 = 0$, $n_3 = 3$, and $n_4 = 5$; a representative plot is given in Fig.\ref{fig:conversion_factor_16} for operator 16. The rest of the multiplicatively renormalizable minus-type operators (10,12, and 14) follow the same format as operator 16.

The plots of Figs.~\ref{fig:conversion_factors_7_8}--\ref{fig:conversion_factor_16} show the real part of the conversion factors as a function of $z/a$. They highlight data points at integer values of $z/a$ ranging from 1 to $L/2=16$, while dashed lines connecting these points display the conversion factors for arbitrary noninteger values of z/a. The value at $z/a=0$ has been excluded from the analysis; indeed, a singular behavior is expected at $z=0$, where the nonlocal operator collapses to a local composite operator with additional contact singularities. The imaginary part of plus-type operators is strictly zero given our choice of renormalization conditions. For minus-type operators the imaginary part is negligible, having a magnitude less than $10^{-5}$. In these plots, we include all possible positive values of $z$ up to half the lattice size, focusing only on the positive directions of the Wilson line. By definition of the plus-type and minus-type operators and the selected RI$'$ renormalization conditions, the conversion factors are symmetric with respect to $z=0$, and therefore, negative values of $z$ are not shown in the plots.

We note here the divergent behavior shown in the plots of `minus-type' operators (Figs.~\ref{fig:conversion_factors_5_6} and \ref{fig:conversion_factor_16}) for noninteger values of $z/a$; this is due to the unavoidable factor of $\sin(z {\bar q}_3)$ in their tree-level expressions (cf. Eqs.~(\ref{eq:RIprime_3}, \ref{eq:RIprime_4}, \ref{eq:RIprime_6}, \ref{eq:RIprime_7}), which renders these expressions non-invertible for some noninteger values of $z/a$). Of course, $z/a$ is necessarily an integer in the lattice definition of the operators, making these divergences inconsequential; however, this behavior points out the necessity for a judicious choice of the renormalization 4-vector scale ${\bar q}$, as mentioned above, so that no divergences will occur at integer values of $z/a$.

\subsection{Lattice Regularization}

We now focus on computing the bare Green's functions, as given by Eq.~(\ref{eq:Green_f_nonlocal}), using lattice regularization. The tree-level Green's functions yield the same result as in dimensional regularization, shown in Eq.~(\ref{eq:Tree_level_green_function}).

The 1-loop computation is considerably more complicated than in dimensional regularization due to the subtleties involved in extracting divergences from lattice integrals. To begin, we write the lattice expressions in the form of a sum of continuum integrals plus additional lattice corrections. Noteworthy, these additional terms, although they have a simple quadratic dependence on the external momentum $q$, are expected to have a nontrivial dependence on $z$ as seen in nonlocal fermion operators \cite{Constantinou2017}. 

Several diagrams (1, 2, 4, 5, 8, 10, and 14), as seen in Fig.~\ref{fig:Feynman_diagrams}, give precisely the same contributions as in DR. This aligns with expectations, considering that these contributions are finite as $\epsilon \rightarrow 0$. Consequently, the limit $a \rightarrow 0$ can be applied right from the beginning, without inducing any lattice corrections. However, we must ensure that we eliminate the overall factor of $1/a^2$, attributed to the presence of the external gluons in the Green's functions, by extracting two powers of the external momentum, $(aq)$.

As an example of the ensuing expressions, we present the one-loop lattice result for diagram 13, which is particularly simple, but includes all types of divergences found in our calculations:
\begin{equation}
    \Lambda^{\text{d13}}_{O_{\mu \nu \rho \sigma}} = \frac{g_0^2 N_c}{16 \pi^2} \left( c_1 + c_2 \, \beta - c_3 \frac{|z|}{a} + 8 \log{\frac{|z|}{a}}\, (2+\beta) \right)  \Lambda_{O_{\mu \nu \rho \sigma}}^{\text{tree}}
    \label{eq:lattice_d13}
\end{equation}
where $c_1=32.24812(2)$, $c_2=14.24059(4)$, and $c_3=79.81936(8)$. Note here the presence of both linear divergence and logarithmic divergence in $a$, features revealed in the nonlocal fermion operators as well. Other diagrams typically have more complicated tensorial structures than the tree level, and also contain a very complex dependence on the momenta of the Green's function, in terms of the integrals over Bessel functions shown in Appendix~\ref{ap:Feynman_parameter_integrals}. The complete expression for $\Lambda^{\rm \overline{MS}}$ can be found in the supplemental file.

The difference between the bare lattice Green’s functions and the $\rm \overline{MS}$-renormalized ones, calculated up to one loop, is as follows:
\begin{align}
    \Lambda_{O_{\mu \nu \rho \sigma}}^{ \rm DR, \overline{\mathrm{MS}}} - \Lambda_{O_{\mu \nu \rho \sigma}}^{\rm LR}= 
    \frac{g^2}{16 \pi^2} \Bigg\{ -\frac{4 \pi^2}{N_c} + N_c  \Bigg[   &\left( \alpha_1 + \log(a^2 \bar{\mu}^2 ) \right) \left( \delta_{\mu 3}+\delta_{\nu 3}+\delta_{\rho 3}+\delta_{\sigma 3} \right) \nonumber \\ 
    &+ \alpha_2 + \alpha_3 \beta + \alpha_4 \frac{\abs{z}}{a} - \frac{\beta}{2} \log(a^2 \bar{\mu}^2 ) \Bigg]  \Bigg\} \Lambda_{O_{\mu \nu \rho \sigma}}^{\text{tree}}
    \label{eq:diff_DR_LR_result}
\end{align}
where $\alpha_1=-8.37940$, $\alpha_2=36.04994$, $\alpha_3=1.38629$, and  $\alpha_4=19.95484$. Despite the extremely complicated momentum dependence and the complex tensorial structure of both the $\rm \overline{MS}$ and the bare lattice Green's functions, their difference (Eq.~\ref{eq:diff_DR_LR_result})is proportional to the tree-level Green's function, indicating multiplicative renormalization without mixing in $\rm \overline{MS}$; the proportionality factor is momentum-independent, as expected. Note that the coefficient $\alpha_4$ in front of the linear divergence has the same value as the corresponding divergent term in the quark nonlocal operators of an arbitrary Wilson line's shape~\cite{Spanoudes:2024kpb}. This is a consequence of the fact that linear divergence arises only from Wilson-line self-energy. As expected, the linear divergent term depends on the length of the Wilson lines and logarithmic divergences arise from the endpoints and contact points of the Wilson lines.

Using the above equation together with Eq.~(\ref{eq:diff_DR_LR}) one can extract the multiplicative renormalization and mixing coefficients in LR using the $\rm \overline{MS}$-scheme. The value found for the of coefficient $\alpha_3$ was expected, since all gauge dependence must disappear in the $\rm \overline{MS}$ scheme for gauge-invariant operators: Indeed, this term will cancel against a similar term in $Z^{\text{LR}, \overline{\rm MS}}_{A}$ in Eq.~(\ref{eq:renorm_factor_gluon_field_lattice}).
For clover-improved Wilson fermions the latter has the value \cite{Gracey2003}:
\begin{align}
    Z^{\text{LR}, \overline{\rm MS}}_{A}= 1+ \frac{g^2}{16 \pi^2} \Bigg\{ &- \frac{2 \pi^2}{N_c} + N_f \left(e^A_1 + e^A_2 c_{SW} + e^A_3 c_{SW}^2 \right) + N_c \left(e^A_4 + e^A_5 \beta \right)  \nonumber \\ 
    &+ \left[ \left( -\frac{5}{3}-\frac{\beta}{2} \right) N_c +\frac{2 N_f}{3} \right] \log( a^2 \bar{\mu}^2 ) \Bigg\}
    \label{eq:renorm_factor_gluon_field_lattice}
\end{align} 
where $e^A_1=-1.05739$, $e^A_2=0.79694$, $e^A_3=-4.71269$, $e^A_4=18.2349$, $e^A_5=1.38629$, and
$c_{SW}$ is the standard clover coefficient \cite{Horsley2004}.

At the one-loop level, the renormalization factors of the operators are found to be diagonal, in both original basis ($O_{\mu \nu \rho \sigma}$) and basis shown in Table~\ref{tb:parity_rotation_symmetries}, as observed in the case of DR. This implies that in the lattice theory at the 1-loop level, the nonlocal gluon operators under investigation are multiplicatively renormalized. By using Eq.~\ref{eq:diff_DR_LR}, one can derive:
\begin{align}
    Z^{\text{LR}, \overline{\rm MS}}_{O_{\mu \nu \rho \sigma}}= 1+ \frac{g^2}{16 \pi^2} \Bigg\{ &\frac{2 \pi^2}{N_c} + N_f \left(e_1 + e_2\, c_{SW} + e_3\, c_{SW}^2 + \frac{2}{3} \log(a^2 \bar{\mu}^2 ) \right) \nonumber \\ 
    &+ N_c \Bigg[e_4 + e_5 \frac{\abs{z}}{a} - \frac{5}{3} \log(a^2 \bar{\mu}^2 ) - \left( e_6 + \log(a^2 \bar{\mu}^2 ) \right) \left( \delta_{\mu 3}+\delta_{\nu 3}+\delta_{\rho 3}+\delta_{\sigma 3} \right) \Bigg]  \Bigg\} 
    \label{eq:renorm_factor_operator_lattice}
\end{align}
where $e_1=e^A_1=-1.05739$, $e_2=e^A_2=0.79694$, $e_3=e^A_3=-4.71269$, $e_4=-17.81504$, $e_5=-\alpha_4=-19.95484$, and $e_6=\alpha_1=-8.37940$. It is worth mentioning that the presence of $c_{SW}$ in $Z^{\text{LR}, \overline{\rm MS}}_{O_{\mu \nu \rho \sigma}}$ is inherited from $Z^{\text{LR}, \overline{\rm MS}}_{A}$. As expected, $Z^{\text{LR}, \overline{\rm MS}}_{O_{\mu \nu \rho \sigma}}$ is gauge independent, and the cancellation of the gauge dependence was numerically confirmed up to $\mathcal{O}(10^{-5})$. This gives an estimate of the accuracy of the numerical loop integration. 

Similarly to Eq.~\ref{eq:renorm_factor_operator_new_basis}, in the basis of Table~\ref{tb:parity_rotation_symmetries},  the matrix $Z^{\rm LR, \rm \overline{MS}}_{ij}$ takes the following diagonal form:
\begin{align}
    Z^{\text{LR}, \overline{\rm MS}}_{ij}= \delta_{ij} \Bigg[ 1+ \frac{g^2}{16 \pi^2} \Bigg\{ &\frac{2 \pi^2}{N_c} + N_f \left(e_1 + e_2\, c_{SW} + e_3\, c_{SW}^2 + \frac{2}{3} \log(a^2 \bar{\mu}^2 ) \right) \nonumber \\ 
    &+ N_c \left(e_4 + e_5 \frac{\abs{z}}{a} - \frac{5}{3} \log(a^2 \bar{\mu}^2 ) - \left( e_6 + \log(a^2 \bar{\mu}^2 ) \right) \omega_{i} \right)  \Bigg\}  \Bigg]
    \label{eq:renorm_factor_operator_lattice_new_basis}
\end{align}
where $\omega_{i}$ is defined by Eq.~\ref{eq:omega}.

Even though the one-loop lattice calculation shows a multiplicative renormalization for all the gluon nonlocal operators under study, we expect that mixing among pairs of operators, as dictated by the symmetries of QCD, will be revealed at higher orders. The absence of mixing at one loop, found in our calculation, provides a valuable input to the nonperturbative studies regarding the size of mixing contributions expected to arise in lattice simulations: Although a multiplicatively renormalizable operator is a better candidate to explore the hadron matrix elements of gluon PDFs, in practice, other operators, which can mix only at higher orders of perturbation theory, can be possible alternatives, if their mixing contributions are small enough compared to statistical errors, and thus, negligible.

\section{Summary}
\label{sec:conclusions}

In this paper, we have studied the renormalization of the gluon nonlocal operators. By analyzing the symmetry properties of these operators, we have identified their mixing pattern under renormalization; some undergo mixing into pairs ($\{1,\ 2\}$, $\{3,\ 4\}$, $\{5,\ 6\}$, $\{7,\ 8\}$ for notation, see Table~\ref{tb:parity_rotation_symmetries}), while others are multiplicatively renormalizable (9-16). We have computed the two-point bare Green’s functions of gluon nonlocal operators using both dimensional and lattice regularization methods. We have evaluated the renormalization factors in the $\rm \overline{MS}$ scheme. At the one-loop level, the renormalization factors for the operators were found to be diagonal, both in the continuum and on the lattice. This implies that in lattice theory, at the 1-loop level, the nonlocal gluon operators undergo multiplicative renormalization. This observation aligns with the pattern revealed by symmetry arguments: mixing is expected to occur, albeit at higher orders of perturbation theory. Additionally, we determined the conversion factors of these operators between the RI$'$ and $\rm \overline{MS}$ renormalization schemes. The RI$'$ scheme was defined to be compatible with the mixing pattern of the operators and be practical for nonperturbative studies. The outcomes of this study are essential for exploring potential paths for investigations of gluon PDFs through lattice QCD. Furthermore, they contribute insights into the renormalization of general gluon nonlocal operators on the lattice, thereby facilitating the development of nonperturbative renormalization prescriptions.

There is a number of possible extensions to this work. One particular direction is the study of higher-order effects beyond the one-loop level. Another extension regards using a number of improved lattice actions and investigating their effect on the renormalization factors. Further, the calculation of additional Green's functions and use of variant renormalization schemes will allow for alternative ways of renormalizing the nonlocal operators, enabling stringent cross-checks when converting to ${\rm \overline{MS}}$\,; one possible variant scheme is a coordinate-space Gauge-Invariant Renormalization scheme \cite{Costa:2021iyv}. This broader investigation can provide a tight control of sources of systematic error, which is essential for nonperturbative studies. 

\begin{acknowledgments}
The project (EXCELLENCE/0421/0025) is implemented under the program of social cohesion ``THALIA 2021-2027" co-funded by the European Union, through Research and Innovation Foundation. G.S. acknowledges funding under the project 3D-nucleon, contract number EXCELLENCE/0421/0043, co-financed by the European Regional Development Fund and the Republic of Cyprus through the Cyprus Research and Innovation Foundation.
\end{acknowledgments}

\appendix

\section{Character Table of Octahedral point group}
\label{ap:character_table}

Table~\ref{tb:characters} provides an overview of the representations of the rotational octahedral point group. Each row corresponds to an irreducible representation, while the columns denote the classes of symmetry operations, including the identity operation ($E$) and rotations ($C_n$) along different axes. For our purposes it is sufficient to focus on classes $C_{2}$ representing $180^{\circ}$ rotations about each of the 3 axes perpendicular to the Wilson line and $C_{4}$ representing $90^{\circ}$ rotations about these axes. 

\begin{table}[h!]
\centering
\begin{tabular}{c|ccccc}
 & $E$ & $8 C_{3}$ & $3 C_{2}=3 C_{4}^{2}$ & $6 C_{2}^{\prime}$ & $6 C_{4}$ \\
\hline
$A_{1}$ & 1 & 1 & 1 & 1 & 1 \\
$A_{2}$ & 1 & 1 & 1 & -1 & -1 \\
$E$ & 2 & -1 & 2 & 0 & 0 \\
$T_{1}$ & 3 & 0 & -1 & -1 & 1 \\
$T_{2}$ & 3 & 0 & -1 & 1 & -1 
\end{tabular}
\caption{Character table of the rotational octahedral point group.}
\label{tb:characters}
\end{table}

\section{Definition of Feynman parameter Integrals}
\label{ap:Feynman_parameter_integrals}

In this appendix we provide a list of Feynman parameter integrals, featured in the expressions of the conversion factors, which lack a closed analytical form. Notably, all the integrals discussed in this context are convergent and their numerical calculation is straightforward.


These integrals depend on both the external momentum 4-vector $q_\nu$ and the length of the Wilson line, $z$.  Within the integrands, we encounter modified Bessel functions of the second kind, denoted as $K_{0}$ and $K_{1}$. To simplify notation, we introduce the parameter $s \equiv \sqrt{q^2(1-x)x}$. All integrals are dimensionless by definition.

\begin{align}
F_{1}(q^2,q_3, z) & =\int_{0}^{1} dx \, e^{-i q_3 x z} \, K_{0}(s |z|)  \label{eq:BesselK_integrals_F1} \\
F_{2}(q^2,q_3, z) & =\int_{0}^{1} dx \, e^{-i q_3 x z} \, K_{0}(s |z|) \, x \\
F_{3}(q^2,q_3, z) & =\int_{0}^{1} dx \, e^{-i q_3 x z} \, K_{0}(s |z|) \, x^2 \\
F_{4}(q^2,q_3, z) & =\int_{0}^{1} dx \, e^{-i q_3 x z} \, K_{0}(s |z|) \, x^3 \\
F_{5}(q^2,q_3, z) & =\int_{0}^{1} dx \, e^{-i q_3 x z} \, K_{0}(s |z|) \, x^4 \\
F_{6}(q^2,q_3, z) & =\int_{0}^{1} dx \, e^{-i q_3 x z} \, K_{0}(s |z|) \, x^5
\end{align}
\begin{align}
F_{7}(q^2,q_3, z) & =\int_{0}^{1} dx \, e^{-i q_3 x z} \, K_{1}(s |z|) \, s|z| \\
F_{8}(q^2,q_3, z) & =\int_{0}^{1} dx \, e^{-i q_3 x z} \, K_{1}(s |z|) \, s|z| \, x \\
F_{9}(q^2,q_3, z) & =\int_{0}^{1} dx \, e^{-i q_3 x z} \, K_{1}(s |z|) \, s|z| \, x^2 
\label{eq:BesselK_integrals_F9}
\end{align}

For `minus-type' operators there appear also double integrals of modified Bessel functions over both $x$ and the parameter $\zeta$; an example is provided below:

\begin{equation}
    \int_{0}^{1} dx \, \int_{0}^{z} d\zeta \, e^{-i q_3 x \zeta} \, K_{0}(s |\zeta|) \, \frac{1}{|z|} \nonumber
\end{equation}

\bibliography{references}

\begin{thebibliography}{67}%
\makeatletter
\providecommand \@ifxundefined [1]{%
 \@ifx{#1\undefined}
}%
\providecommand \@ifnum [1]{%
 \ifnum #1\expandafter \@firstoftwo
 \else \expandafter \@secondoftwo
 \fi
}%
\providecommand \@ifx [1]{%
 \ifx #1\expandafter \@firstoftwo
 \else \expandafter \@secondoftwo
 \fi
}%
\providecommand \natexlab [1]{#1}%
\providecommand \enquote  [1]{``#1''}%
\providecommand \bibnamefont  [1]{#1}%
\providecommand \bibfnamefont [1]{#1}%
\providecommand \citenamefont [1]{#1}%
\providecommand \href@noop [0]{\@secondoftwo}%
\providecommand \href [0]{\begingroup \@sanitize@url \@href}%
\providecommand \@href[1]{\@@startlink{#1}\@@href}%
\providecommand \@@href[1]{\endgroup#1\@@endlink}%
\providecommand \@sanitize@url [0]{\catcode `\\12\catcode `\$12\catcode `\&12\catcode `\#12\catcode `\^12\catcode `\_12\catcode `\%12\relax}%
\providecommand \@@startlink[1]{}%
\providecommand \@@endlink[0]{}%
\providecommand \url  [0]{\begingroup\@sanitize@url \@url }%
\providecommand \@url [1]{\endgroup\@href {#1}{\urlprefix }}%
\providecommand \urlprefix  [0]{URL }%
\providecommand \Eprint [0]{\href }%
\providecommand \doibase [0]{https://doi.org/}%
\providecommand \selectlanguage [0]{\@gobble}%
\providecommand \bibinfo  [0]{\@secondoftwo}%
\providecommand \bibfield  [0]{\@secondoftwo}%
\providecommand \translation [1]{[#1]}%
\providecommand \BibitemOpen [0]{}%
\providecommand \bibitemStop [0]{}%
\providecommand \bibitemNoStop [0]{.\EOS\space}%
\providecommand \EOS [0]{\spacefactor3000\relax}%
\providecommand \BibitemShut  [1]{\csname bibitem#1\endcsname}%
\let\auto@bib@innerbib\@empty
\bibitem [{\citenamefont {Collins}\ \emph {et~al.}(1989)\citenamefont {Collins}, \citenamefont {Soper},\ and\ \citenamefont {Sterman}}]{Collins:1989gx}%
  \BibitemOpen
  \bibfield  {author} {\bibinfo {author} {\bibfnamefont {J.~C.}\ \bibnamefont {Collins}}, \bibinfo {author} {\bibfnamefont {D.~E.}\ \bibnamefont {Soper}},\ and\ \bibinfo {author} {\bibfnamefont {G.~F.}\ \bibnamefont {Sterman}},\ }\bibfield  {title} {\bibinfo {title} {{Factorization of Hard Processes in QCD}},\ }\href {https://doi.org/10.1142/9789814503266_0001} {\bibfield  {journal} {\bibinfo  {journal} {Adv. Ser. Direct. High Energy Phys.}\ }\textbf {\bibinfo {volume} {5}},\ \bibinfo {pages} {1} (\bibinfo {year} {1989})},\ \Eprint {https://arxiv.org/abs/hep-ph/0409313} {arXiv:hep-ph/0409313} \BibitemShut {NoStop}%
\bibitem [{\citenamefont {Ethier}\ \emph {et~al.}(2017)\citenamefont {Ethier}, \citenamefont {Sato},\ and\ \citenamefont {Melnitchouk}}]{Ethier:2017zbq}%
  \BibitemOpen
  \bibfield  {author} {\bibinfo {author} {\bibfnamefont {J.~J.}\ \bibnamefont {Ethier}}, \bibinfo {author} {\bibfnamefont {N.}~\bibnamefont {Sato}},\ and\ \bibinfo {author} {\bibfnamefont {W.}~\bibnamefont {Melnitchouk}},\ }\bibfield  {title} {\bibinfo {title} {{First simultaneous extraction of spin-dependent parton distributions and fragmentation functions from a global QCD analysis}},\ }\href {https://doi.org/10.1103/PhysRevLett.119.132001} {\bibfield  {journal} {\bibinfo  {journal} {Phys. Rev. Lett.}\ }\textbf {\bibinfo {volume} {119}},\ \bibinfo {pages} {132001} (\bibinfo {year} {2017})},\ \Eprint {https://arxiv.org/abs/1705.05889} {arXiv:1705.05889 [hep-ph]} \BibitemShut {NoStop}%
\bibitem [{\citenamefont {Nocera}\ \emph {et~al.}(2014)\citenamefont {Nocera}, \citenamefont {Ball}, \citenamefont {Forte}, \citenamefont {Ridolfi},\ and\ \citenamefont {Rojo}}]{Nocera:2014gqa}%
  \BibitemOpen
  \bibfield  {author} {\bibinfo {author} {\bibfnamefont {E.~R.}\ \bibnamefont {Nocera}}, \bibinfo {author} {\bibfnamefont {R.~D.}\ \bibnamefont {Ball}}, \bibinfo {author} {\bibfnamefont {S.}~\bibnamefont {Forte}}, \bibinfo {author} {\bibfnamefont {G.}~\bibnamefont {Ridolfi}},\ and\ \bibinfo {author} {\bibfnamefont {J.}~\bibnamefont {Rojo}} (\bibinfo {collaboration} {NNPDF}),\ }\bibfield  {title} {\bibinfo {title} {{A first unbiased global determination of polarized PDFs and their uncertainties}},\ }\href {https://doi.org/10.1016/j.nuclphysb.2014.08.008} {\bibfield  {journal} {\bibinfo  {journal} {Nucl. Phys. B}\ }\textbf {\bibinfo {volume} {887}},\ \bibinfo {pages} {276} (\bibinfo {year} {2014})},\ \Eprint {https://arxiv.org/abs/1406.5539} {arXiv:1406.5539 [hep-ph]} \BibitemShut {NoStop}%
\bibitem [{\citenamefont {de~Florian}\ \emph {et~al.}(2009)\citenamefont {de~Florian}, \citenamefont {Sassot}, \citenamefont {Stratmann},\ and\ \citenamefont {Vogelsang}}]{deFlorian:2009vb}%
  \BibitemOpen
  \bibfield  {author} {\bibinfo {author} {\bibfnamefont {D.}~\bibnamefont {de~Florian}}, \bibinfo {author} {\bibfnamefont {R.}~\bibnamefont {Sassot}}, \bibinfo {author} {\bibfnamefont {M.}~\bibnamefont {Stratmann}},\ and\ \bibinfo {author} {\bibfnamefont {W.}~\bibnamefont {Vogelsang}},\ }\bibfield  {title} {\bibinfo {title} {{Extraction of Spin-Dependent Parton Densities and Their Uncertainties}},\ }\href {https://doi.org/10.1103/PhysRevD.80.034030} {\bibfield  {journal} {\bibinfo  {journal} {Phys. Rev. D}\ }\textbf {\bibinfo {volume} {80}},\ \bibinfo {pages} {034030} (\bibinfo {year} {2009})},\ \Eprint {https://arxiv.org/abs/0904.3821} {arXiv:0904.3821 [hep-ph]} \BibitemShut {NoStop}%
\bibitem [{\citenamefont {Accardi}\ \emph {et~al.}(2016)\citenamefont {Accardi}, \citenamefont {Brady}, \citenamefont {Melnitchouk}, \citenamefont {Owens},\ and\ \citenamefont {Sato}}]{Accardi:2016qay}%
  \BibitemOpen
  \bibfield  {author} {\bibinfo {author} {\bibfnamefont {A.}~\bibnamefont {Accardi}}, \bibinfo {author} {\bibfnamefont {L.~T.}\ \bibnamefont {Brady}}, \bibinfo {author} {\bibfnamefont {W.}~\bibnamefont {Melnitchouk}}, \bibinfo {author} {\bibfnamefont {J.~F.}\ \bibnamefont {Owens}},\ and\ \bibinfo {author} {\bibfnamefont {N.}~\bibnamefont {Sato}},\ }\bibfield  {title} {\bibinfo {title} {{Constraints on large-$x$ parton distributions from new weak boson production and deep-inelastic scattering data}},\ }\href {https://doi.org/10.1103/PhysRevD.93.114017} {\bibfield  {journal} {\bibinfo  {journal} {Phys. Rev. D}\ }\textbf {\bibinfo {volume} {93}},\ \bibinfo {pages} {114017} (\bibinfo {year} {2016})},\ \Eprint {https://arxiv.org/abs/1602.03154} {arXiv:1602.03154 [hep-ph]} \BibitemShut {NoStop}%
\bibitem [{\citenamefont {Ball}\ \emph {et~al.}(2017)\citenamefont {Ball} \emph {et~al.}}]{NNPDF:2017mvq}%
  \BibitemOpen
  \bibfield  {author} {\bibinfo {author} {\bibfnamefont {R.~D.}\ \bibnamefont {Ball}} \emph {et~al.} (\bibinfo {collaboration} {NNPDF}),\ }\bibfield  {title} {\bibinfo {title} {{Parton distributions from high-precision collider data}},\ }\href {https://doi.org/10.1140/epjc/s10052-017-5199-5} {\bibfield  {journal} {\bibinfo  {journal} {Eur. Phys. J. C}\ }\textbf {\bibinfo {volume} {77}},\ \bibinfo {pages} {663} (\bibinfo {year} {2017})},\ \Eprint {https://arxiv.org/abs/1706.00428} {arXiv:1706.00428 [hep-ph]} \BibitemShut {NoStop}%
\bibitem [{\citenamefont {Alekhin}\ \emph {et~al.}(2017)\citenamefont {Alekhin}, \citenamefont {Bl\"umlein}, \citenamefont {Moch},\ and\ \citenamefont {Placakyte}}]{Alekhin:2017kpj}%
  \BibitemOpen
  \bibfield  {author} {\bibinfo {author} {\bibfnamefont {S.}~\bibnamefont {Alekhin}}, \bibinfo {author} {\bibfnamefont {J.}~\bibnamefont {Bl\"umlein}}, \bibinfo {author} {\bibfnamefont {S.}~\bibnamefont {Moch}},\ and\ \bibinfo {author} {\bibfnamefont {R.}~\bibnamefont {Placakyte}},\ }\bibfield  {title} {\bibinfo {title} {{Parton distribution functions, $\alpha_s$, and heavy-quark masses for LHC Run II}},\ }\href {https://doi.org/10.1103/PhysRevD.96.014011} {\bibfield  {journal} {\bibinfo  {journal} {Phys. Rev. D}\ }\textbf {\bibinfo {volume} {96}},\ \bibinfo {pages} {014011} (\bibinfo {year} {2017})},\ \Eprint {https://arxiv.org/abs/1701.05838} {arXiv:1701.05838 [hep-ph]} \BibitemShut {NoStop}%
\bibitem [{\citenamefont {Dolgov}\ \emph {et~al.}(2002)\citenamefont {Dolgov} \emph {et~al.}}]{LHPC:2002xzk}%
  \BibitemOpen
  \bibfield  {author} {\bibinfo {author} {\bibfnamefont {D.}~\bibnamefont {Dolgov}} \emph {et~al.} (\bibinfo {collaboration} {LHPC, TXL}),\ }\bibfield  {title} {\bibinfo {title} {{Moments of nucleon light cone quark distributions calculated in full lattice QCD}},\ }\href {https://doi.org/10.1103/PhysRevD.66.034506} {\bibfield  {journal} {\bibinfo  {journal} {Phys. Rev. D}\ }\textbf {\bibinfo {volume} {66}},\ \bibinfo {pages} {034506} (\bibinfo {year} {2002})},\ \Eprint {https://arxiv.org/abs/hep-lat/0201021} {arXiv:hep-lat/0201021} \BibitemShut {NoStop}%
\bibitem [{\citenamefont {Dolgov}\ \emph {et~al.}(2001)\citenamefont {Dolgov} \emph {et~al.}}]{Dolgov:2000ca}%
  \BibitemOpen
  \bibfield  {author} {\bibinfo {author} {\bibfnamefont {D.}~\bibnamefont {Dolgov}} \emph {et~al.},\ }\bibfield  {title} {\bibinfo {title} {{Moments of structure functions in full QCD}},\ }\href {https://doi.org/10.1016/S0920-5632(01)00943-4} {\bibfield  {journal} {\bibinfo  {journal} {Nucl. Phys. B Proc. Suppl.}\ }\textbf {\bibinfo {volume} {94}},\ \bibinfo {pages} {303} (\bibinfo {year} {2001})},\ \Eprint {https://arxiv.org/abs/hep-lat/0011010} {arXiv:hep-lat/0011010} \BibitemShut {NoStop}%
\bibitem [{\citenamefont {Dreher}\ \emph {et~al.}(2003)\citenamefont {Dreher} \emph {et~al.}}]{SESAM:2002zfm}%
  \BibitemOpen
  \bibfield  {author} {\bibinfo {author} {\bibfnamefont {P.}~\bibnamefont {Dreher}} \emph {et~al.} (\bibinfo {collaboration} {SESAM, LHPC}),\ }\bibfield  {title} {\bibinfo {title} {{Continuum extrapolation of moments of nucleon quark distributions in full QCD}},\ }\href {https://doi.org/10.1016/S0920-5632(03)01564-0} {\bibfield  {journal} {\bibinfo  {journal} {Nucl. Phys. B Proc. Suppl.}\ }\textbf {\bibinfo {volume} {119}},\ \bibinfo {pages} {392} (\bibinfo {year} {2003})},\ \Eprint {https://arxiv.org/abs/hep-lat/0211021} {arXiv:hep-lat/0211021} \BibitemShut {NoStop}%
\bibitem [{\citenamefont {Gockeler}\ \emph {et~al.}(2003)\citenamefont {Gockeler}, \citenamefont {Horsley}, \citenamefont {Pleiter}, \citenamefont {Rakow}, \citenamefont {Schafer},\ and\ \citenamefont {Schierholz}}]{Gockeler:2002ek}%
  \BibitemOpen
  \bibfield  {author} {\bibinfo {author} {\bibfnamefont {M.}~\bibnamefont {Gockeler}}, \bibinfo {author} {\bibfnamefont {R.}~\bibnamefont {Horsley}}, \bibinfo {author} {\bibfnamefont {D.}~\bibnamefont {Pleiter}}, \bibinfo {author} {\bibfnamefont {P.~E.~L.}\ \bibnamefont {Rakow}}, \bibinfo {author} {\bibfnamefont {A.}~\bibnamefont {Schafer}},\ and\ \bibinfo {author} {\bibfnamefont {G.}~\bibnamefont {Schierholz}},\ }\bibfield  {title} {\bibinfo {title} {{Calculation of moments of structure functions}},\ }\href {https://doi.org/10.1016/S0920-5632(03)01490-7} {\bibfield  {journal} {\bibinfo  {journal} {Nucl. Phys. B Proc. Suppl.}\ }\textbf {\bibinfo {volume} {119}},\ \bibinfo {pages} {32} (\bibinfo {year} {2003})},\ \Eprint {https://arxiv.org/abs/hep-lat/0209160} {arXiv:hep-lat/0209160} \BibitemShut {NoStop}%
\bibitem [{\citenamefont {Cichy}\ and\ \citenamefont {Constantinou}(2019)}]{Cichy:2018mum}%
  \BibitemOpen
  \bibfield  {author} {\bibinfo {author} {\bibfnamefont {K.}~\bibnamefont {Cichy}}\ and\ \bibinfo {author} {\bibfnamefont {M.}~\bibnamefont {Constantinou}},\ }\bibfield  {title} {\bibinfo {title} {{A guide to light-cone PDFs from Lattice QCD: an overview of approaches, techniques and results}},\ }\href {https://doi.org/10.1155/2019/3036904} {\bibfield  {journal} {\bibinfo  {journal} {Adv. High Energy Phys.}\ }\textbf {\bibinfo {volume} {2019}},\ \bibinfo {pages} {3036904} (\bibinfo {year} {2019})},\ \Eprint {https://arxiv.org/abs/1811.07248} {arXiv:1811.07248 [hep-lat]} \BibitemShut {NoStop}%
\bibitem [{\citenamefont {Ji}\ \emph {et~al.}(2021)\citenamefont {Ji}, \citenamefont {Liu}, \citenamefont {Liu}, \citenamefont {Zhang},\ and\ \citenamefont {Zhao}}]{Ji:2020ect}%
  \BibitemOpen
  \bibfield  {author} {\bibinfo {author} {\bibfnamefont {X.}~\bibnamefont {Ji}}, \bibinfo {author} {\bibfnamefont {Y.-S.}\ \bibnamefont {Liu}}, \bibinfo {author} {\bibfnamefont {Y.}~\bibnamefont {Liu}}, \bibinfo {author} {\bibfnamefont {J.-H.}\ \bibnamefont {Zhang}},\ and\ \bibinfo {author} {\bibfnamefont {Y.}~\bibnamefont {Zhao}},\ }\bibfield  {title} {\bibinfo {title} {{Large-momentum effective theory}},\ }\href {https://doi.org/10.1103/RevModPhys.93.035005} {\bibfield  {journal} {\bibinfo  {journal} {Rev. Mod. Phys.}\ }\textbf {\bibinfo {volume} {93}},\ \bibinfo {pages} {035005} (\bibinfo {year} {2021})},\ \Eprint {https://arxiv.org/abs/2004.03543} {arXiv:2004.03543 [hep-ph]} \BibitemShut {NoStop}%
\bibitem [{\citenamefont {Constantinou}(2021)}]{Constantinou:2020pek}%
  \BibitemOpen
  \bibfield  {author} {\bibinfo {author} {\bibfnamefont {M.}~\bibnamefont {Constantinou}},\ }\bibfield  {title} {\bibinfo {title} {{The x-dependence of hadronic parton distributions: A review on the progress of lattice QCD}},\ }\href {https://doi.org/10.1140/epja/s10050-021-00353-7} {\bibfield  {journal} {\bibinfo  {journal} {Eur. Phys. J. A}\ }\textbf {\bibinfo {volume} {57}},\ \bibinfo {pages} {77} (\bibinfo {year} {2021})},\ \Eprint {https://arxiv.org/abs/2010.02445} {arXiv:2010.02445 [hep-lat]} \BibitemShut {NoStop}%
\bibitem [{\citenamefont {Cichy}(2022{\natexlab{a}})}]{Cichy:2021lih}%
  \BibitemOpen
  \bibfield  {author} {\bibinfo {author} {\bibfnamefont {K.}~\bibnamefont {Cichy}},\ }\bibfield  {title} {\bibinfo {title} {{Progress in $x$-dependent partonic distributions from lattice QCD}},\ }\href {https://doi.org/10.22323/1.396.0017} {\bibfield  {journal} {\bibinfo  {journal} {PoS}\ }\textbf {\bibinfo {volume} {LATTICE2021}},\ \bibinfo {pages} {017} (\bibinfo {year} {2022}{\natexlab{a}})},\ \Eprint {https://arxiv.org/abs/2110.07440} {arXiv:2110.07440 [hep-lat]} \BibitemShut {NoStop}%
\bibitem [{\citenamefont {Cichy}(2022{\natexlab{b}})}]{Cichy:2021ewm}%
  \BibitemOpen
  \bibfield  {author} {\bibinfo {author} {\bibfnamefont {K.}~\bibnamefont {Cichy}},\ }\bibfield  {title} {\bibinfo {title} {{Overview of lattice calculations of the x-dependence of PDFs, GPDs and TMDs}},\ }\href {https://doi.org/10.1051/epjconf/202225801005} {\bibfield  {journal} {\bibinfo  {journal} {EPJ Web Conf.}\ }\textbf {\bibinfo {volume} {258}},\ \bibinfo {pages} {01005} (\bibinfo {year} {2022}{\natexlab{b}})},\ \Eprint {https://arxiv.org/abs/2111.04552} {arXiv:2111.04552 [hep-lat]} \BibitemShut {NoStop}%
\bibitem [{\citenamefont {Ji}(2013)}]{Ji2013}%
  \BibitemOpen
  \bibfield  {author} {\bibinfo {author} {\bibfnamefont {X.}~\bibnamefont {Ji}},\ }\bibfield  {title} {\bibinfo {title} {{Parton Physics on a Euclidean Lattice}},\ }\href {https://doi.org/10.1103/PhysRevLett.110.262002} {\bibfield  {journal} {\bibinfo  {journal} {Phys. Rev. Lett.}\ }\textbf {\bibinfo {volume} {110}},\ \bibinfo {pages} {262002} (\bibinfo {year} {2013})},\ \Eprint {https://arxiv.org/abs/1305.1539} {arXiv:1305.1539 [hep-ph]} \BibitemShut {NoStop}%
\bibitem [{\citenamefont {Ji}(2014)}]{Ji2014}%
  \BibitemOpen
  \bibfield  {author} {\bibinfo {author} {\bibfnamefont {X.}~\bibnamefont {Ji}},\ }\bibfield  {title} {\bibinfo {title} {{Parton Physics from Large-Momentum Effective Field Theory}},\ }\href {https://doi.org/10.1007/s11433-014-5492-3} {\bibfield  {journal} {\bibinfo  {journal} {Sci. China Phys. Mech. Astron.}\ }\textbf {\bibinfo {volume} {57}},\ \bibinfo {pages} {1407} (\bibinfo {year} {2014})},\ \Eprint {https://arxiv.org/abs/1404.6680} {arXiv:1404.6680 [hep-ph]} \BibitemShut {NoStop}%
\bibitem [{\citenamefont {Ma}\ and\ \citenamefont {Qiu}(2018)}]{Ma2014}%
  \BibitemOpen
  \bibfield  {author} {\bibinfo {author} {\bibfnamefont {Y.-Q.}\ \bibnamefont {Ma}}\ and\ \bibinfo {author} {\bibfnamefont {J.-W.}\ \bibnamefont {Qiu}},\ }\bibfield  {title} {\bibinfo {title} {{Extracting Parton Distribution Functions from Lattice QCD Calculations}},\ }\href {https://doi.org/10.1103/PhysRevD.98.074021} {\bibfield  {journal} {\bibinfo  {journal} {Phys. Rev. D}\ }\textbf {\bibinfo {volume} {98}},\ \bibinfo {pages} {074021} (\bibinfo {year} {2018})},\ \Eprint {https://arxiv.org/abs/1404.6860} {arXiv:1404.6860 [hep-ph]} \BibitemShut {NoStop}%
\bibitem [{\citenamefont {Wang}\ \emph {et~al.}(2019{\natexlab{a}})\citenamefont {Wang}, \citenamefont {Zhang}, \citenamefont {Zhao},\ and\ \citenamefont {Zhu}}]{Wang:2019tgg}%
  \BibitemOpen
  \bibfield  {author} {\bibinfo {author} {\bibfnamefont {W.}~\bibnamefont {Wang}}, \bibinfo {author} {\bibfnamefont {J.-H.}\ \bibnamefont {Zhang}}, \bibinfo {author} {\bibfnamefont {S.}~\bibnamefont {Zhao}},\ and\ \bibinfo {author} {\bibfnamefont {R.}~\bibnamefont {Zhu}},\ }\bibfield  {title} {\bibinfo {title} {{Complete matching for quasidistribution functions in large momentum effective theory}},\ }\href {https://doi.org/10.1103/PhysRevD.100.074509} {\bibfield  {journal} {\bibinfo  {journal} {Phys. Rev. D}\ }\textbf {\bibinfo {volume} {100}},\ \bibinfo {pages} {074509} (\bibinfo {year} {2019}{\natexlab{a}})},\ \Eprint {https://arxiv.org/abs/1904.00978} {arXiv:1904.00978 [hep-ph]} \BibitemShut {NoStop}%
\bibitem [{\citenamefont {Wang}\ \emph {et~al.}(2018)\citenamefont {Wang}, \citenamefont {Zhao},\ and\ \citenamefont {Zhu}}]{Wang:2017qyg}%
  \BibitemOpen
  \bibfield  {author} {\bibinfo {author} {\bibfnamefont {W.}~\bibnamefont {Wang}}, \bibinfo {author} {\bibfnamefont {S.}~\bibnamefont {Zhao}},\ and\ \bibinfo {author} {\bibfnamefont {R.}~\bibnamefont {Zhu}},\ }\bibfield  {title} {\bibinfo {title} {{Gluon quasidistribution function at one loop}},\ }\href {https://doi.org/10.1140/epjc/s10052-018-5617-3} {\bibfield  {journal} {\bibinfo  {journal} {Eur. Phys. J. C}\ }\textbf {\bibinfo {volume} {78}},\ \bibinfo {pages} {147} (\bibinfo {year} {2018})},\ \Eprint {https://arxiv.org/abs/1708.02458} {arXiv:1708.02458 [hep-ph]} \BibitemShut {NoStop}%
\bibitem [{\citenamefont {Chen}\ \emph {et~al.}(2020)\citenamefont {Chen}, \citenamefont {Wang},\ and\ \citenamefont {Zhu}}]{Chen:2020arf}%
  \BibitemOpen
  \bibfield  {author} {\bibinfo {author} {\bibfnamefont {L.-B.}\ \bibnamefont {Chen}}, \bibinfo {author} {\bibfnamefont {W.}~\bibnamefont {Wang}},\ and\ \bibinfo {author} {\bibfnamefont {R.}~\bibnamefont {Zhu}},\ }\bibfield  {title} {\bibinfo {title} {{Quasi parton distribution functions at NNLO: flavor non-diagonal quark contributions}},\ }\href {https://doi.org/10.1103/PhysRevD.102.011503} {\bibfield  {journal} {\bibinfo  {journal} {Phys. Rev. D}\ }\textbf {\bibinfo {volume} {102}},\ \bibinfo {pages} {011503} (\bibinfo {year} {2020})},\ \Eprint {https://arxiv.org/abs/2005.13757} {arXiv:2005.13757 [hep-ph]} \BibitemShut {NoStop}%
\bibitem [{\citenamefont {Li}\ \emph {et~al.}(2021)\citenamefont {Li}, \citenamefont {Ma},\ and\ \citenamefont {Qiu}}]{Li:2020xml}%
  \BibitemOpen
  \bibfield  {author} {\bibinfo {author} {\bibfnamefont {Z.-Y.}\ \bibnamefont {Li}}, \bibinfo {author} {\bibfnamefont {Y.-Q.}\ \bibnamefont {Ma}},\ and\ \bibinfo {author} {\bibfnamefont {J.-W.}\ \bibnamefont {Qiu}},\ }\bibfield  {title} {\bibinfo {title} {{Extraction of Next-to-Next-to-Leading-Order Parton Distribution Functions from Lattice QCD Calculations}},\ }\href {https://doi.org/10.1103/PhysRevLett.126.072001} {\bibfield  {journal} {\bibinfo  {journal} {Phys. Rev. Lett.}\ }\textbf {\bibinfo {volume} {126}},\ \bibinfo {pages} {072001} (\bibinfo {year} {2021})},\ \Eprint {https://arxiv.org/abs/2006.12370} {arXiv:2006.12370 [hep-ph]} \BibitemShut {NoStop}%
\bibitem [{\citenamefont {Alexandrou}\ \emph {et~al.}(2017{\natexlab{a}})\citenamefont {Alexandrou}, \citenamefont {Constantinou}, \citenamefont {Hadjiyiannakou}, \citenamefont {Jansen}, \citenamefont {Kallidonis}, \citenamefont {Koutsou}, \citenamefont {Vaquero Avil\'es-Casco},\ and\ \citenamefont {Wiese}}]{Alexandrou:2017oeh}%
  \BibitemOpen
  \bibfield  {author} {\bibinfo {author} {\bibfnamefont {C.}~\bibnamefont {Alexandrou}}, \bibinfo {author} {\bibfnamefont {M.}~\bibnamefont {Constantinou}}, \bibinfo {author} {\bibfnamefont {K.}~\bibnamefont {Hadjiyiannakou}}, \bibinfo {author} {\bibfnamefont {K.}~\bibnamefont {Jansen}}, \bibinfo {author} {\bibfnamefont {C.}~\bibnamefont {Kallidonis}}, \bibinfo {author} {\bibfnamefont {G.}~\bibnamefont {Koutsou}}, \bibinfo {author} {\bibfnamefont {A.}~\bibnamefont {Vaquero Avil\'es-Casco}},\ and\ \bibinfo {author} {\bibfnamefont {C.}~\bibnamefont {Wiese}},\ }\bibfield  {title} {\bibinfo {title} {{Nucleon Spin and Momentum Decomposition Using Lattice QCD Simulations}},\ }\href {https://doi.org/10.1103/PhysRevLett.119.142002} {\bibfield  {journal} {\bibinfo  {journal} {Phys. Rev. Lett.}\ }\textbf {\bibinfo {volume} {119}},\ \bibinfo {pages} {142002} (\bibinfo {year} {2017}{\natexlab{a}})},\ \Eprint {https://arxiv.org/abs/1706.02973} {arXiv:1706.02973 [hep-lat]} \BibitemShut {NoStop}%
\bibitem [{\citenamefont {Alexandrou}\ \emph {et~al.}(2020)\citenamefont {Alexandrou}, \citenamefont {Bacchio}, \citenamefont {Constantinou}, \citenamefont {Finkenrath}, \citenamefont {Hadjiyiannakou}, \citenamefont {Jansen}, \citenamefont {Koutsou}, \citenamefont {Panagopoulos},\ and\ \citenamefont {Spanoudes}}]{Alexandrou:2020sml}%
  \BibitemOpen
  \bibfield  {author} {\bibinfo {author} {\bibfnamefont {C.}~\bibnamefont {Alexandrou}}, \bibinfo {author} {\bibfnamefont {S.}~\bibnamefont {Bacchio}}, \bibinfo {author} {\bibfnamefont {M.}~\bibnamefont {Constantinou}}, \bibinfo {author} {\bibfnamefont {J.}~\bibnamefont {Finkenrath}}, \bibinfo {author} {\bibfnamefont {K.}~\bibnamefont {Hadjiyiannakou}}, \bibinfo {author} {\bibfnamefont {K.}~\bibnamefont {Jansen}}, \bibinfo {author} {\bibfnamefont {G.}~\bibnamefont {Koutsou}}, \bibinfo {author} {\bibfnamefont {H.}~\bibnamefont {Panagopoulos}},\ and\ \bibinfo {author} {\bibfnamefont {G.}~\bibnamefont {Spanoudes}},\ }\bibfield  {title} {\bibinfo {title} {{Complete flavor decomposition of the spin and momentum fraction of the proton using lattice QCD simulations at physical pion mass}},\ }\href {https://doi.org/10.1103/PhysRevD.101.094513} {\bibfield  {journal} {\bibinfo  {journal} {Phys. Rev. D}\ }\textbf {\bibinfo {volume} {101}},\ \bibinfo {pages} {094513} (\bibinfo {year} {2020})},\ \Eprint
  {https://arxiv.org/abs/2003.08486} {arXiv:2003.08486 [hep-lat]} \BibitemShut {NoStop}%
\bibitem [{\citenamefont {Yang}\ \emph {et~al.}(2018)\citenamefont {Yang}, \citenamefont {Gong}, \citenamefont {Liang}, \citenamefont {Lin}, \citenamefont {Liu}, \citenamefont {Pefkou},\ and\ \citenamefont {Shanahan}}]{Yang:2018bft}%
  \BibitemOpen
  \bibfield  {author} {\bibinfo {author} {\bibfnamefont {Y.-B.}\ \bibnamefont {Yang}}, \bibinfo {author} {\bibfnamefont {M.}~\bibnamefont {Gong}}, \bibinfo {author} {\bibfnamefont {J.}~\bibnamefont {Liang}}, \bibinfo {author} {\bibfnamefont {H.-W.}\ \bibnamefont {Lin}}, \bibinfo {author} {\bibfnamefont {K.-F.}\ \bibnamefont {Liu}}, \bibinfo {author} {\bibfnamefont {D.}~\bibnamefont {Pefkou}},\ and\ \bibinfo {author} {\bibfnamefont {P.}~\bibnamefont {Shanahan}},\ }\bibfield  {title} {\bibinfo {title} {{Nonperturbatively renormalized glue momentum fraction at the physical pion mass from lattice QCD}},\ }\href {https://doi.org/10.1103/PhysRevD.98.074506} {\bibfield  {journal} {\bibinfo  {journal} {Phys. Rev. D}\ }\textbf {\bibinfo {volume} {98}},\ \bibinfo {pages} {074506} (\bibinfo {year} {2018})},\ \Eprint {https://arxiv.org/abs/1805.00531} {arXiv:1805.00531 [hep-lat]} \BibitemShut {NoStop}%
\bibitem [{\citenamefont {Alekhin}\ \emph {et~al.}(2014)\citenamefont {Alekhin}, \citenamefont {Bl\"umlein},\ and\ \citenamefont {Moch}}]{Alekhin2014}%
  \BibitemOpen
  \bibfield  {author} {\bibinfo {author} {\bibfnamefont {S.}~\bibnamefont {Alekhin}}, \bibinfo {author} {\bibfnamefont {J.}~\bibnamefont {Bl\"umlein}},\ and\ \bibinfo {author} {\bibfnamefont {S.}~\bibnamefont {Moch}},\ }\bibfield  {title} {\bibinfo {title} {The abm parton distributions tuned to lhc data},\ }\href {https://doi.org/10.1103/PhysRevD.89.054028} {\bibfield  {journal} {\bibinfo  {journal} {Phys. Rev. D}\ }\textbf {\bibinfo {volume} {89}},\ \bibinfo {pages} {054028} (\bibinfo {year} {2014})}\BibitemShut {NoStop}%
\bibitem [{\citenamefont {Butterworth}\ \emph {et~al.}(2016)\citenamefont {Butterworth} \emph {et~al.}}]{Butterworth:2015oua}%
  \BibitemOpen
  \bibfield  {author} {\bibinfo {author} {\bibfnamefont {J.}~\bibnamefont {Butterworth}} \emph {et~al.},\ }\bibfield  {title} {\bibinfo {title} {{PDF4LHC recommendations for LHC Run II}},\ }\href {https://doi.org/10.1088/0954-3899/43/2/023001} {\bibfield  {journal} {\bibinfo  {journal} {J. Phys. G}\ }\textbf {\bibinfo {volume} {43}},\ \bibinfo {pages} {023001} (\bibinfo {year} {2016})},\ \Eprint {https://arxiv.org/abs/1510.03865} {arXiv:1510.03865 [hep-ph]} \BibitemShut {NoStop}%
\bibitem [{\citenamefont {Sato}\ \emph {et~al.}(2020)\citenamefont {Sato}, \citenamefont {Andres}, \citenamefont {Ethier},\ and\ \citenamefont {Melnitchouk}}]{Sato:2019yez}%
  \BibitemOpen
  \bibfield  {author} {\bibinfo {author} {\bibfnamefont {N.}~\bibnamefont {Sato}}, \bibinfo {author} {\bibfnamefont {C.}~\bibnamefont {Andres}}, \bibinfo {author} {\bibfnamefont {J.~J.}\ \bibnamefont {Ethier}},\ and\ \bibinfo {author} {\bibfnamefont {W.}~\bibnamefont {Melnitchouk}} (\bibinfo {collaboration} {JAM}),\ }\bibfield  {title} {\bibinfo {title} {{Strange quark suppression from a simultaneous Monte Carlo analysis of parton distributions and fragmentation functions}},\ }\href {https://doi.org/10.1103/PhysRevD.101.074020} {\bibfield  {journal} {\bibinfo  {journal} {Phys. Rev. D}\ }\textbf {\bibinfo {volume} {101}},\ \bibinfo {pages} {074020} (\bibinfo {year} {2020})},\ \Eprint {https://arxiv.org/abs/1905.03788} {arXiv:1905.03788 [hep-ph]} \BibitemShut {NoStop}%
\bibitem [{\citenamefont {Hou}\ \emph {et~al.}(2017)\citenamefont {Hou}, \citenamefont {Dulat}, \citenamefont {Gao}, \citenamefont {Guzzi}, \citenamefont {Huston}, \citenamefont {Nadolsky}, \citenamefont {Pumplin}, \citenamefont {Schmidt}, \citenamefont {Stump},\ and\ \citenamefont {Yuan}}]{Hou:2016nqm}%
  \BibitemOpen
  \bibfield  {author} {\bibinfo {author} {\bibfnamefont {T.-J.}\ \bibnamefont {Hou}}, \bibinfo {author} {\bibfnamefont {S.}~\bibnamefont {Dulat}}, \bibinfo {author} {\bibfnamefont {J.}~\bibnamefont {Gao}}, \bibinfo {author} {\bibfnamefont {M.}~\bibnamefont {Guzzi}}, \bibinfo {author} {\bibfnamefont {J.}~\bibnamefont {Huston}}, \bibinfo {author} {\bibfnamefont {P.}~\bibnamefont {Nadolsky}}, \bibinfo {author} {\bibfnamefont {J.}~\bibnamefont {Pumplin}}, \bibinfo {author} {\bibfnamefont {C.}~\bibnamefont {Schmidt}}, \bibinfo {author} {\bibfnamefont {D.}~\bibnamefont {Stump}},\ and\ \bibinfo {author} {\bibfnamefont {C.~P.}\ \bibnamefont {Yuan}},\ }\bibfield  {title} {\bibinfo {title} {{CTEQ-TEA parton distribution functions and HERA Run I and II combined data}},\ }\href {https://doi.org/10.1103/PhysRevD.95.034003} {\bibfield  {journal} {\bibinfo  {journal} {Phys. Rev. D}\ }\textbf {\bibinfo {volume} {95}},\ \bibinfo {pages} {034003} (\bibinfo {year} {2017})},\ \Eprint {https://arxiv.org/abs/1609.07968}
  {arXiv:1609.07968 [hep-ph]} \BibitemShut {NoStop}%
\bibitem [{\citenamefont {Harland-Lang}\ \emph {et~al.}(2015)\citenamefont {Harland-Lang}, \citenamefont {Martin}, \citenamefont {Motylinski},\ and\ \citenamefont {Thorne}}]{Harland-Lang:2014zoa}%
  \BibitemOpen
  \bibfield  {author} {\bibinfo {author} {\bibfnamefont {L.~A.}\ \bibnamefont {Harland-Lang}}, \bibinfo {author} {\bibfnamefont {A.~D.}\ \bibnamefont {Martin}}, \bibinfo {author} {\bibfnamefont {P.}~\bibnamefont {Motylinski}},\ and\ \bibinfo {author} {\bibfnamefont {R.~S.}\ \bibnamefont {Thorne}},\ }\bibfield  {title} {\bibinfo {title} {{Parton distributions in the LHC era: MMHT 2014 PDFs}},\ }\href {https://doi.org/10.1140/epjc/s10052-015-3397-6} {\bibfield  {journal} {\bibinfo  {journal} {Eur. Phys. J. C}\ }\textbf {\bibinfo {volume} {75}},\ \bibinfo {pages} {204} (\bibinfo {year} {2015})},\ \Eprint {https://arxiv.org/abs/1412.3989} {arXiv:1412.3989 [hep-ph]} \BibitemShut {NoStop}%
\bibitem [{\citenamefont {Moffat}\ \emph {et~al.}(2021)\citenamefont {Moffat}, \citenamefont {Melnitchouk}, \citenamefont {Rogers},\ and\ \citenamefont {Sato}}]{Moffat2021}%
  \BibitemOpen
  \bibfield  {author} {\bibinfo {author} {\bibfnamefont {E.}~\bibnamefont {Moffat}}, \bibinfo {author} {\bibfnamefont {W.}~\bibnamefont {Melnitchouk}}, \bibinfo {author} {\bibfnamefont {T.~C.}\ \bibnamefont {Rogers}},\ and\ \bibinfo {author} {\bibfnamefont {N.}~\bibnamefont {Sato}} (\bibinfo {collaboration} {Jefferson Lab Angular Momentum (JAM) Collaboration}),\ }\bibfield  {title} {\bibinfo {title} {Simultaneous monte carlo analysis of parton densities and fragmentation functions},\ }\href {https://doi.org/10.1103/PhysRevD.104.016015} {\bibfield  {journal} {\bibinfo  {journal} {Phys. Rev. D}\ }\textbf {\bibinfo {volume} {104}},\ \bibinfo {pages} {016015} (\bibinfo {year} {2021})}\BibitemShut {NoStop}%
\bibitem [{\citenamefont {Collins}(2023)}]{Collins:1984xc}%
  \BibitemOpen
  \bibfield  {author} {\bibinfo {author} {\bibfnamefont {J.~C.}\ \bibnamefont {Collins}},\ }\href {https://doi.org/10.1017/9781009401807} {\emph {\bibinfo {title} {{Renormalization}}}},\ \bibinfo {series} {Cambridge Monographs on Mathematical Physics}, Vol.~\bibinfo {volume} {26}\ (\bibinfo  {publisher} {Cambridge University Press},\ \bibinfo {address} {Cambridge},\ \bibinfo {year} {2023})\BibitemShut {NoStop}%
\bibitem [{\citenamefont {Wang}\ \emph {et~al.}(2019{\natexlab{b}})\citenamefont {Wang}, \citenamefont {Zhang}, \citenamefont {Zhao},\ and\ \citenamefont {Zhu}}]{Wang2019}%
  \BibitemOpen
  \bibfield  {author} {\bibinfo {author} {\bibfnamefont {W.}~\bibnamefont {Wang}}, \bibinfo {author} {\bibfnamefont {J.-H.}\ \bibnamefont {Zhang}}, \bibinfo {author} {\bibfnamefont {S.}~\bibnamefont {Zhao}},\ and\ \bibinfo {author} {\bibfnamefont {R.}~\bibnamefont {Zhu}},\ }\bibfield  {title} {\bibinfo {title} {{Complete matching for quasidistribution functions in large momentum effective theory}},\ }\href {https://doi.org/10.1103/PhysRevD.100.074509} {\bibfield  {journal} {\bibinfo  {journal} {Phys. Rev. D}\ }\textbf {\bibinfo {volume} {100}},\ \bibinfo {pages} {074509} (\bibinfo {year} {2019}{\natexlab{b}})},\ \Eprint {https://arxiv.org/abs/1904.00978} {arXiv:1904.00978 [hep-ph]} \BibitemShut {NoStop}%
\bibitem [{\citenamefont {Zhang}\ \emph {et~al.}(2019{\natexlab{a}})\citenamefont {Zhang}, \citenamefont {Ji}, \citenamefont {Sch\"afer}, \citenamefont {Wang},\ and\ \citenamefont {Zhao}}]{Zhang:2018diq}%
  \BibitemOpen
  \bibfield  {author} {\bibinfo {author} {\bibfnamefont {J.-H.}\ \bibnamefont {Zhang}}, \bibinfo {author} {\bibfnamefont {X.}~\bibnamefont {Ji}}, \bibinfo {author} {\bibfnamefont {A.}~\bibnamefont {Sch\"afer}}, \bibinfo {author} {\bibfnamefont {W.}~\bibnamefont {Wang}},\ and\ \bibinfo {author} {\bibfnamefont {S.}~\bibnamefont {Zhao}},\ }\bibfield  {title} {\bibinfo {title} {{Accessing Gluon Parton Distributions in Large Momentum Effective Theory}},\ }\href {https://doi.org/10.1103/PhysRevLett.122.142001} {\bibfield  {journal} {\bibinfo  {journal} {Phys. Rev. Lett.}\ }\textbf {\bibinfo {volume} {122}},\ \bibinfo {pages} {142001} (\bibinfo {year} {2019}{\natexlab{a}})},\ \Eprint {https://arxiv.org/abs/1808.10824} {arXiv:1808.10824 [hep-ph]} \BibitemShut {NoStop}%
\bibitem [{\citenamefont {Fan}\ \emph {et~al.}(2018)\citenamefont {Fan}, \citenamefont {Yang}, \citenamefont {Anthony}, \citenamefont {Lin},\ and\ \citenamefont {Liu}}]{Fan:2018dxu}%
  \BibitemOpen
  \bibfield  {author} {\bibinfo {author} {\bibfnamefont {Z.-Y.}\ \bibnamefont {Fan}}, \bibinfo {author} {\bibfnamefont {Y.-B.}\ \bibnamefont {Yang}}, \bibinfo {author} {\bibfnamefont {A.}~\bibnamefont {Anthony}}, \bibinfo {author} {\bibfnamefont {H.-W.}\ \bibnamefont {Lin}},\ and\ \bibinfo {author} {\bibfnamefont {K.-F.}\ \bibnamefont {Liu}},\ }\bibfield  {title} {\bibinfo {title} {{Gluon Quasi-Parton-Distribution Functions from Lattice QCD}},\ }\href {https://doi.org/10.1103/PhysRevLett.121.242001} {\bibfield  {journal} {\bibinfo  {journal} {Phys. Rev. Lett.}\ }\textbf {\bibinfo {volume} {121}},\ \bibinfo {pages} {242001} (\bibinfo {year} {2018})},\ \Eprint {https://arxiv.org/abs/1808.02077} {arXiv:1808.02077 [hep-lat]} \BibitemShut {NoStop}%
\bibitem [{\citenamefont {Balitsky}\ \emph {et~al.}(2020)\citenamefont {Balitsky}, \citenamefont {Morris},\ and\ \citenamefont {Radyushkin}}]{Balitsky:2019krf}%
  \BibitemOpen
  \bibfield  {author} {\bibinfo {author} {\bibfnamefont {I.}~\bibnamefont {Balitsky}}, \bibinfo {author} {\bibfnamefont {W.}~\bibnamefont {Morris}},\ and\ \bibinfo {author} {\bibfnamefont {A.}~\bibnamefont {Radyushkin}},\ }\bibfield  {title} {\bibinfo {title} {{Gluon Pseudo-Distributions at Short Distances: Forward Case}},\ }\href {https://doi.org/10.1016/j.physletb.2020.135621} {\bibfield  {journal} {\bibinfo  {journal} {Phys. Lett. B}\ }\textbf {\bibinfo {volume} {808}},\ \bibinfo {pages} {135621} (\bibinfo {year} {2020})},\ \Eprint {https://arxiv.org/abs/1910.13963} {arXiv:1910.13963 [hep-ph]} \BibitemShut {NoStop}%
\bibitem [{\citenamefont {Fan}\ \emph {et~al.}(2021)\citenamefont {Fan}, \citenamefont {Zhang},\ and\ \citenamefont {Lin}}]{Fan:2020cpa}%
  \BibitemOpen
  \bibfield  {author} {\bibinfo {author} {\bibfnamefont {Z.}~\bibnamefont {Fan}}, \bibinfo {author} {\bibfnamefont {R.}~\bibnamefont {Zhang}},\ and\ \bibinfo {author} {\bibfnamefont {H.-W.}\ \bibnamefont {Lin}},\ }\bibfield  {title} {\bibinfo {title} {{Nucleon gluon distribution function from 2 + 1 + 1-flavor lattice QCD}},\ }\href {https://doi.org/10.1142/S0217751X21500809} {\bibfield  {journal} {\bibinfo  {journal} {Int. J. Mod. Phys. A}\ }\textbf {\bibinfo {volume} {36}},\ \bibinfo {pages} {2150080} (\bibinfo {year} {2021})},\ \Eprint {https://arxiv.org/abs/2007.16113} {arXiv:2007.16113 [hep-lat]} \BibitemShut {NoStop}%
\bibitem [{\citenamefont {Fan}\ and\ \citenamefont {Lin}(2021)}]{Fan:2021bcr}%
  \BibitemOpen
  \bibfield  {author} {\bibinfo {author} {\bibfnamefont {Z.}~\bibnamefont {Fan}}\ and\ \bibinfo {author} {\bibfnamefont {H.-W.}\ \bibnamefont {Lin}},\ }\bibfield  {title} {\bibinfo {title} {{Gluon parton distribution of the pion from lattice QCD}},\ }\href {https://doi.org/10.1016/j.physletb.2021.136778} {\bibfield  {journal} {\bibinfo  {journal} {Phys. Lett. B}\ }\textbf {\bibinfo {volume} {823}},\ \bibinfo {pages} {136778} (\bibinfo {year} {2021})},\ \Eprint {https://arxiv.org/abs/2104.06372} {arXiv:2104.06372 [hep-lat]} \BibitemShut {NoStop}%
\bibitem [{\citenamefont {Salas-Chavira}\ \emph {et~al.}(2022)\citenamefont {Salas-Chavira}, \citenamefont {Fan},\ and\ \citenamefont {Lin}}]{Salas-Chavira:2021wui}%
  \BibitemOpen
  \bibfield  {author} {\bibinfo {author} {\bibfnamefont {A.}~\bibnamefont {Salas-Chavira}}, \bibinfo {author} {\bibfnamefont {Z.}~\bibnamefont {Fan}},\ and\ \bibinfo {author} {\bibfnamefont {H.-W.}\ \bibnamefont {Lin}},\ }\bibfield  {title} {\bibinfo {title} {{First glimpse into the kaon gluon parton distribution using lattice QCD}},\ }\href {https://doi.org/10.1103/PhysRevD.106.094510} {\bibfield  {journal} {\bibinfo  {journal} {Phys. Rev. D}\ }\textbf {\bibinfo {volume} {106}},\ \bibinfo {pages} {094510} (\bibinfo {year} {2022})},\ \Eprint {https://arxiv.org/abs/2112.03124} {arXiv:2112.03124 [hep-lat]} \BibitemShut {NoStop}%
\bibitem [{\citenamefont {Khan}\ \emph {et~al.}(2021)\citenamefont {Khan} \emph {et~al.}}]{HadStruc:2021wmh}%
  \BibitemOpen
  \bibfield  {author} {\bibinfo {author} {\bibfnamefont {T.}~\bibnamefont {Khan}} \emph {et~al.} (\bibinfo {collaboration} {HadStruc}),\ }\bibfield  {title} {\bibinfo {title} {{Unpolarized gluon distribution in the nucleon from lattice quantum chromodynamics}},\ }\href {https://doi.org/10.1103/PhysRevD.104.094516} {\bibfield  {journal} {\bibinfo  {journal} {Phys. Rev. D}\ }\textbf {\bibinfo {volume} {104}},\ \bibinfo {pages} {094516} (\bibinfo {year} {2021})},\ \Eprint {https://arxiv.org/abs/2107.08960} {arXiv:2107.08960 [hep-lat]} \BibitemShut {NoStop}%
\bibitem [{\citenamefont {Egerer}\ \emph {et~al.}(2022)\citenamefont {Egerer} \emph {et~al.}}]{HadStruc:2022yaw}%
  \BibitemOpen
  \bibfield  {author} {\bibinfo {author} {\bibfnamefont {C.}~\bibnamefont {Egerer}} \emph {et~al.} (\bibinfo {collaboration} {HadStruc}),\ }\bibfield  {title} {\bibinfo {title} {{Toward the determination of the gluon helicity distribution in the nucleon from lattice quantum chromodynamics}},\ }\href {https://doi.org/10.1103/PhysRevD.106.094511} {\bibfield  {journal} {\bibinfo  {journal} {Phys. Rev. D}\ }\textbf {\bibinfo {volume} {106}},\ \bibinfo {pages} {094511} (\bibinfo {year} {2022})},\ \Eprint {https://arxiv.org/abs/2207.08733} {arXiv:2207.08733 [hep-lat]} \BibitemShut {NoStop}%
\bibitem [{\citenamefont {Khan}\ \emph {et~al.}(2023)\citenamefont {Khan}, \citenamefont {Liu},\ and\ \citenamefont {Sufian}}]{Khan:2022vot}%
  \BibitemOpen
  \bibfield  {author} {\bibinfo {author} {\bibfnamefont {T.}~\bibnamefont {Khan}}, \bibinfo {author} {\bibfnamefont {T.}~\bibnamefont {Liu}},\ and\ \bibinfo {author} {\bibfnamefont {R.~S.}\ \bibnamefont {Sufian}},\ }\bibfield  {title} {\bibinfo {title} {{Gluon helicity in the nucleon from lattice QCD and machine learning}},\ }\href {https://doi.org/10.1103/PhysRevD.108.074502} {\bibfield  {journal} {\bibinfo  {journal} {Phys. Rev. D}\ }\textbf {\bibinfo {volume} {108}},\ \bibinfo {pages} {074502} (\bibinfo {year} {2023})},\ \Eprint {https://arxiv.org/abs/2211.15587} {arXiv:2211.15587 [hep-lat]} \BibitemShut {NoStop}%
\bibitem [{\citenamefont {Fan}\ \emph {et~al.}(2023)\citenamefont {Fan}, \citenamefont {Good},\ and\ \citenamefont {Lin}}]{Fan:2022kcb}%
  \BibitemOpen
  \bibfield  {author} {\bibinfo {author} {\bibfnamefont {Z.}~\bibnamefont {Fan}}, \bibinfo {author} {\bibfnamefont {W.}~\bibnamefont {Good}},\ and\ \bibinfo {author} {\bibfnamefont {H.-W.}\ \bibnamefont {Lin}},\ }\bibfield  {title} {\bibinfo {title} {{Gluon parton distribution of the nucleon from (2+1+1)-flavor lattice QCD in the physical-continuum limit}},\ }\href {https://doi.org/10.1103/PhysRevD.108.014508} {\bibfield  {journal} {\bibinfo  {journal} {Phys. Rev. D}\ }\textbf {\bibinfo {volume} {108}},\ \bibinfo {pages} {014508} (\bibinfo {year} {2023})},\ \Eprint {https://arxiv.org/abs/2210.09985} {arXiv:2210.09985 [hep-lat]} \BibitemShut {NoStop}%
\bibitem [{\citenamefont {Delmar}\ \emph {et~al.}(2023)\citenamefont {Delmar}, \citenamefont {Alexandrou}, \citenamefont {Cichy}, \citenamefont {Constantinou},\ and\ \citenamefont {Hadjiyiannakou}}]{Delmar:2023agv}%
  \BibitemOpen
  \bibfield  {author} {\bibinfo {author} {\bibfnamefont {J.}~\bibnamefont {Delmar}}, \bibinfo {author} {\bibfnamefont {C.}~\bibnamefont {Alexandrou}}, \bibinfo {author} {\bibfnamefont {K.}~\bibnamefont {Cichy}}, \bibinfo {author} {\bibfnamefont {M.}~\bibnamefont {Constantinou}},\ and\ \bibinfo {author} {\bibfnamefont {K.}~\bibnamefont {Hadjiyiannakou}},\ }\bibfield  {title} {\bibinfo {title} {{Gluon PDF of the proton using twisted mass fermions}},\ }\href {https://doi.org/10.1103/PhysRevD.108.094515} {\bibfield  {journal} {\bibinfo  {journal} {Phys. Rev. D}\ }\textbf {\bibinfo {volume} {108}},\ \bibinfo {pages} {094515} (\bibinfo {year} {2023})},\ \Eprint {https://arxiv.org/abs/2310.01389} {arXiv:2310.01389 [hep-lat]} \BibitemShut {NoStop}%
\bibitem [{\citenamefont {Constantinou}\ and\ \citenamefont {Panagopoulos}(2017)}]{Constantinou2017}%
  \BibitemOpen
  \bibfield  {author} {\bibinfo {author} {\bibfnamefont {M.}~\bibnamefont {Constantinou}}\ and\ \bibinfo {author} {\bibfnamefont {H.}~\bibnamefont {Panagopoulos}},\ }\bibfield  {title} {\bibinfo {title} {{Perturbative renormalization of quasi-parton distribution functions}},\ }\href {https://doi.org/10.1103/PhysRevD.96.054506} {\bibfield  {journal} {\bibinfo  {journal} {Phys. Rev. D}\ }\textbf {\bibinfo {volume} {96}},\ \bibinfo {pages} {054506} (\bibinfo {year} {2017})},\ \Eprint {https://arxiv.org/abs/1705.11193} {arXiv:1705.11193 [hep-lat]} \BibitemShut {NoStop}%
\bibitem [{\citenamefont {Alexandrou}\ \emph {et~al.}(2017{\natexlab{b}})\citenamefont {Alexandrou}, \citenamefont {Cichy}, \citenamefont {Constantinou}, \citenamefont {Hadjiyiannakou}, \citenamefont {Jansen}, \citenamefont {Panagopoulos},\ and\ \citenamefont {Steffens}}]{Alexandrou2017}%
  \BibitemOpen
  \bibfield  {author} {\bibinfo {author} {\bibfnamefont {C.}~\bibnamefont {Alexandrou}}, \bibinfo {author} {\bibfnamefont {K.}~\bibnamefont {Cichy}}, \bibinfo {author} {\bibfnamefont {M.}~\bibnamefont {Constantinou}}, \bibinfo {author} {\bibfnamefont {K.}~\bibnamefont {Hadjiyiannakou}}, \bibinfo {author} {\bibfnamefont {K.}~\bibnamefont {Jansen}}, \bibinfo {author} {\bibfnamefont {H.}~\bibnamefont {Panagopoulos}},\ and\ \bibinfo {author} {\bibfnamefont {F.}~\bibnamefont {Steffens}},\ }\bibfield  {title} {\bibinfo {title} {{A complete non-perturbative renormalization prescription for quasi-PDFs}},\ }\href {https://doi.org/10.1016/j.nuclphysb.2017.08.012} {\bibfield  {journal} {\bibinfo  {journal} {Nucl. Phys. B}\ }\textbf {\bibinfo {volume} {923}},\ \bibinfo {pages} {394} (\bibinfo {year} {2017}{\natexlab{b}})},\ \Eprint {https://arxiv.org/abs/1706.00265} {arXiv:1706.00265 [hep-lat]} \BibitemShut {NoStop}%
\bibitem [{\citenamefont {Zhang}\ \emph {et~al.}(2019{\natexlab{b}})\citenamefont {Zhang}, \citenamefont {Ji}, \citenamefont {Sch\"afer}, \citenamefont {Wang},\ and\ \citenamefont {Zhao}}]{Zhang2018}%
  \BibitemOpen
  \bibfield  {author} {\bibinfo {author} {\bibfnamefont {J.-H.}\ \bibnamefont {Zhang}}, \bibinfo {author} {\bibfnamefont {X.}~\bibnamefont {Ji}}, \bibinfo {author} {\bibfnamefont {A.}~\bibnamefont {Sch\"afer}}, \bibinfo {author} {\bibfnamefont {W.}~\bibnamefont {Wang}},\ and\ \bibinfo {author} {\bibfnamefont {S.}~\bibnamefont {Zhao}},\ }\bibfield  {title} {\bibinfo {title} {{Accessing Gluon Parton Distributions in Large Momentum Effective Theory}},\ }\href {https://doi.org/10.1103/PhysRevLett.122.142001} {\bibfield  {journal} {\bibinfo  {journal} {Phys. Rev. Lett.}\ }\textbf {\bibinfo {volume} {122}},\ \bibinfo {pages} {142001} (\bibinfo {year} {2019}{\natexlab{b}})},\ \Eprint {https://arxiv.org/abs/1808.10824} {arXiv:1808.10824 [hep-ph]} \BibitemShut {NoStop}%
\bibitem [{\citenamefont {Dorn}\ \emph {et~al.}(1983)\citenamefont {Dorn}, \citenamefont {Robaschik},\ and\ \citenamefont {Wieczorek}}]{Dorn:1981wa}%
  \BibitemOpen
  \bibfield  {author} {\bibinfo {author} {\bibfnamefont {H.}~\bibnamefont {Dorn}}, \bibinfo {author} {\bibfnamefont {D.}~\bibnamefont {Robaschik}},\ and\ \bibinfo {author} {\bibfnamefont {E.}~\bibnamefont {Wieczorek}},\ }\bibfield  {title} {\bibinfo {title} {{RENORMALIZATION AND SHORT DISTANCE PROPERTIES OF GAUGE INVARIANT GLUONIUM AND HADRON OPERATORS}},\ }\href {https://doi.org/10.1002/andp.19834950208} {\bibfield  {journal} {\bibinfo  {journal} {Annalen Phys.}\ }\textbf {\bibinfo {volume} {40}},\ \bibinfo {pages} {166} (\bibinfo {year} {1983})}\BibitemShut {NoStop}%
\bibitem [{\citenamefont {Wang}\ and\ \citenamefont {Zhao}(2018)}]{Wang:2017eel}%
  \BibitemOpen
  \bibfield  {author} {\bibinfo {author} {\bibfnamefont {W.}~\bibnamefont {Wang}}\ and\ \bibinfo {author} {\bibfnamefont {S.}~\bibnamefont {Zhao}},\ }\bibfield  {title} {\bibinfo {title} {{On the power divergence in quasi gluon distribution function}},\ }\href {https://doi.org/10.1007/JHEP05(2018)142} {\bibfield  {journal} {\bibinfo  {journal} {JHEP}\ }\textbf {\bibinfo {volume} {05}},\ \bibinfo {pages} {142}},\ \Eprint {https://arxiv.org/abs/1712.09247} {arXiv:1712.09247 [hep-ph]} \BibitemShut {NoStop}%
\bibitem [{\citenamefont {Braun}\ \emph {et~al.}(2020)\citenamefont {Braun}, \citenamefont {Chetyrkin},\ and\ \citenamefont {Kniehl}}]{Braun:2020ymy}%
  \BibitemOpen
  \bibfield  {author} {\bibinfo {author} {\bibfnamefont {V.~M.}\ \bibnamefont {Braun}}, \bibinfo {author} {\bibfnamefont {K.~G.}\ \bibnamefont {Chetyrkin}},\ and\ \bibinfo {author} {\bibfnamefont {B.~A.}\ \bibnamefont {Kniehl}},\ }\bibfield  {title} {\bibinfo {title} {{Renormalization of parton quasi-distributions beyond the leading order: spacelike vs. timelike}},\ }\href {https://doi.org/10.1007/JHEP07(2020)161} {\bibfield  {journal} {\bibinfo  {journal} {JHEP}\ }\textbf {\bibinfo {volume} {07}},\ \bibinfo {pages} {161}},\ \Eprint {https://arxiv.org/abs/2004.01043} {arXiv:2004.01043 [hep-ph]} \BibitemShut {NoStop}%
\bibitem [{\citenamefont {Sheikholeslami}\ and\ \citenamefont {Wohlert}(1985)}]{Sheikholeslami1985}%
  \BibitemOpen
  \bibfield  {author} {\bibinfo {author} {\bibfnamefont {B.}~\bibnamefont {Sheikholeslami}}\ and\ \bibinfo {author} {\bibfnamefont {R.}~\bibnamefont {Wohlert}},\ }\bibfield  {title} {\bibinfo {title} {Improved continuum limit lattice action for qcd with wilson fermions},\ }\href {https://doi.org/https://doi.org/10.1016/0550-3213(85)90002-1} {\bibfield  {journal} {\bibinfo  {journal} {Nuclear Physics B}\ }\textbf {\bibinfo {volume} {259}},\ \bibinfo {pages} {572} (\bibinfo {year} {1985})}\BibitemShut {NoStop}%
\bibitem [{\citenamefont {Gracey}(2003)}]{Gracey2003}%
  \BibitemOpen
  \bibfield  {author} {\bibinfo {author} {\bibfnamefont {J.~A.}\ \bibnamefont {Gracey}},\ }\bibfield  {title} {\bibinfo {title} {{Three loop anomalous dimension of nonsinglet quark currents in the RI-prime scheme}},\ }\href {https://doi.org/10.1016/S0550-3213(03)00335-3} {\bibfield  {journal} {\bibinfo  {journal} {Nucl. Phys. B}\ }\textbf {\bibinfo {volume} {662}},\ \bibinfo {pages} {247} (\bibinfo {year} {2003})},\ \Eprint {https://arxiv.org/abs/hep-ph/0304113} {arXiv:hep-ph/0304113} \BibitemShut {NoStop}%
\bibitem [{\citenamefont {Dorn}(1986)}]{Dorn:1986dt}%
  \BibitemOpen
  \bibfield  {author} {\bibinfo {author} {\bibfnamefont {H.}~\bibnamefont {Dorn}},\ }\bibfield  {title} {\bibinfo {title} {{Renormalization of Path Ordered Phase Factors and Related Hadron Operators in Gauge Field Theories}},\ }\href {https://doi.org/10.1002/prop.19860340104} {\bibfield  {journal} {\bibinfo  {journal} {Fortsch. Phys.}\ }\textbf {\bibinfo {volume} {34}},\ \bibinfo {pages} {11} (\bibinfo {year} {1986})}\BibitemShut {NoStop}%
\bibitem [{\citenamefont {Dotsenko}\ and\ \citenamefont {Vergeles}(1980)}]{Dotsenko:1979wb}%
  \BibitemOpen
  \bibfield  {author} {\bibinfo {author} {\bibfnamefont {V.~S.}\ \bibnamefont {Dotsenko}}\ and\ \bibinfo {author} {\bibfnamefont {S.~N.}\ \bibnamefont {Vergeles}},\ }\bibfield  {title} {\bibinfo {title} {{Renormalizability of Phase Factors in the Nonabelian Gauge Theory}},\ }\href {https://doi.org/10.1016/0550-3213(80)90103-0} {\bibfield  {journal} {\bibinfo  {journal} {Nucl. Phys. B}\ }\textbf {\bibinfo {volume} {169}},\ \bibinfo {pages} {527} (\bibinfo {year} {1980})}\BibitemShut {NoStop}%
\bibitem [{\citenamefont {Brandt}\ \emph {et~al.}(1981)\citenamefont {Brandt}, \citenamefont {Neri},\ and\ \citenamefont {Sato}}]{Brandt:1981kf}%
  \BibitemOpen
  \bibfield  {author} {\bibinfo {author} {\bibfnamefont {R.~A.}\ \bibnamefont {Brandt}}, \bibinfo {author} {\bibfnamefont {F.}~\bibnamefont {Neri}},\ and\ \bibinfo {author} {\bibfnamefont {M.-a.}\ \bibnamefont {Sato}},\ }\bibfield  {title} {\bibinfo {title} {{Renormalization of Loop Functions for All Loops}},\ }\href {https://doi.org/10.1103/PhysRevD.24.879} {\bibfield  {journal} {\bibinfo  {journal} {Phys. Rev. D}\ }\textbf {\bibinfo {volume} {24}},\ \bibinfo {pages} {879} (\bibinfo {year} {1981})}\BibitemShut {NoStop}%
\bibitem [{\citenamefont {Constantinou}\ \emph {et~al.}(2009)\citenamefont {Constantinou}, \citenamefont {Lubicz}, \citenamefont {Panagopoulos},\ and\ \citenamefont {Stylianou}}]{Constantinou:2009tr}%
  \BibitemOpen
  \bibfield  {author} {\bibinfo {author} {\bibfnamefont {M.}~\bibnamefont {Constantinou}}, \bibinfo {author} {\bibfnamefont {V.}~\bibnamefont {Lubicz}}, \bibinfo {author} {\bibfnamefont {H.}~\bibnamefont {Panagopoulos}},\ and\ \bibinfo {author} {\bibfnamefont {F.}~\bibnamefont {Stylianou}},\ }\bibfield  {title} {\bibinfo {title} {{O(a**2) corrections to the one-loop propagator and bilinears of clover fermions with Symanzik improved gluons}},\ }\href {https://doi.org/10.1088/1126-6708/2009/10/064} {\bibfield  {journal} {\bibinfo  {journal} {JHEP}\ }\textbf {\bibinfo {volume} {10}},\ \bibinfo {pages} {064}},\ \Eprint {https://arxiv.org/abs/0907.0381} {arXiv:0907.0381 [hep-lat]} \BibitemShut {NoStop}%
\bibitem [{\citenamefont {Constantinou}\ \emph {et~al.}(2013)\citenamefont {Constantinou}, \citenamefont {Costa}, \citenamefont {G\"ockeler}, \citenamefont {Horsley}, \citenamefont {Panagopoulos}, \citenamefont {Perlt}, \citenamefont {Rakow}, \citenamefont {Schierholz},\ and\ \citenamefont {Schiller}}]{Constantinou:2013ada}%
  \BibitemOpen
  \bibfield  {author} {\bibinfo {author} {\bibfnamefont {M.}~\bibnamefont {Constantinou}}, \bibinfo {author} {\bibfnamefont {M.}~\bibnamefont {Costa}}, \bibinfo {author} {\bibfnamefont {M.}~\bibnamefont {G\"ockeler}}, \bibinfo {author} {\bibfnamefont {R.}~\bibnamefont {Horsley}}, \bibinfo {author} {\bibfnamefont {H.}~\bibnamefont {Panagopoulos}}, \bibinfo {author} {\bibfnamefont {H.}~\bibnamefont {Perlt}}, \bibinfo {author} {\bibfnamefont {P.~E.~L.}\ \bibnamefont {Rakow}}, \bibinfo {author} {\bibfnamefont {G.}~\bibnamefont {Schierholz}},\ and\ \bibinfo {author} {\bibfnamefont {A.}~\bibnamefont {Schiller}},\ }\bibfield  {title} {\bibinfo {title} {{Perturbatively improving regularization-invariant momentum scheme renormalization constants}},\ }\href {https://doi.org/10.1103/PhysRevD.87.096019} {\bibfield  {journal} {\bibinfo  {journal} {Phys. Rev. D}\ }\textbf {\bibinfo {volume} {87}},\ \bibinfo {pages} {096019} (\bibinfo {year} {2013})},\ \Eprint {https://arxiv.org/abs/1303.6776} {arXiv:1303.6776 [hep-lat]}
  \BibitemShut {NoStop}%
\bibitem [{\citenamefont {Alexandrou}\ \emph {et~al.}(2017{\natexlab{c}})\citenamefont {Alexandrou}, \citenamefont {Constantinou},\ and\ \citenamefont {Panagopoulos}}]{Alexandrou:2015sea}%
  \BibitemOpen
  \bibfield  {author} {\bibinfo {author} {\bibfnamefont {C.}~\bibnamefont {Alexandrou}}, \bibinfo {author} {\bibfnamefont {M.}~\bibnamefont {Constantinou}},\ and\ \bibinfo {author} {\bibfnamefont {H.}~\bibnamefont {Panagopoulos}} (\bibinfo {collaboration} {ETM}),\ }\bibfield  {title} {\bibinfo {title} {{Renormalization functions for Nf=2 and Nf=4 twisted mass fermions}},\ }\href {https://doi.org/10.1103/PhysRevD.95.034505} {\bibfield  {journal} {\bibinfo  {journal} {Phys. Rev. D}\ }\textbf {\bibinfo {volume} {95}},\ \bibinfo {pages} {034505} (\bibinfo {year} {2017}{\natexlab{c}})},\ \Eprint {https://arxiv.org/abs/1509.00213} {arXiv:1509.00213 [hep-lat]} \BibitemShut {NoStop}%
\bibitem [{\citenamefont {Constantinou}\ and\ \citenamefont {Panagopoulos}(2023)}]{Constantinou:2022aij}%
  \BibitemOpen
  \bibfield  {author} {\bibinfo {author} {\bibfnamefont {M.}~\bibnamefont {Constantinou}}\ and\ \bibinfo {author} {\bibfnamefont {H.}~\bibnamefont {Panagopoulos}},\ }\bibfield  {title} {\bibinfo {title} {{Improved renormalization scheme for nonlocal operators}},\ }\href {https://doi.org/10.1103/PhysRevD.107.014503} {\bibfield  {journal} {\bibinfo  {journal} {Phys. Rev. D}\ }\textbf {\bibinfo {volume} {107}},\ \bibinfo {pages} {014503} (\bibinfo {year} {2023})},\ \Eprint {https://arxiv.org/abs/2207.09977} {arXiv:2207.09977 [hep-lat]} \BibitemShut {NoStop}%
\bibitem [{\citenamefont {Joglekar}\ and\ \citenamefont {Lee}(1976)}]{Joglekar:1975nu}%
  \BibitemOpen
  \bibfield  {author} {\bibinfo {author} {\bibfnamefont {S.~D.}\ \bibnamefont {Joglekar}}\ and\ \bibinfo {author} {\bibfnamefont {B.~W.}\ \bibnamefont {Lee}},\ }\bibfield  {title} {\bibinfo {title} {{General Theory of Renormalization of Gauge Invariant Operators}},\ }\href {https://doi.org/10.1016/0003-4916(76)90225-6} {\bibfield  {journal} {\bibinfo  {journal} {Annals Phys.}\ }\textbf {\bibinfo {volume} {97}},\ \bibinfo {pages} {160} (\bibinfo {year} {1976})}\BibitemShut {NoStop}%
\bibitem [{\citenamefont {Rothe}(2012)}]{Rothe:1992nt}%
  \BibitemOpen
  \bibfield  {author} {\bibinfo {author} {\bibfnamefont {H.~J.}\ \bibnamefont {Rothe}},\ }\href {https://doi.org/10.1142/8229} {\emph {\bibinfo {title} {{Lattice Gauge Theories : An Introduction (Fourth Edition)}}}},\ Vol.~\bibinfo {volume} {43}\ (\bibinfo  {publisher} {World Scientific Publishing Company},\ \bibinfo {year} {2012})\BibitemShut {NoStop}%
\bibitem [{\citenamefont {Gattringer}\ and\ \citenamefont {Lang}(2010)}]{Gattringer:2010zz}%
  \BibitemOpen
  \bibfield  {author} {\bibinfo {author} {\bibfnamefont {C.}~\bibnamefont {Gattringer}}\ and\ \bibinfo {author} {\bibfnamefont {C.~B.}\ \bibnamefont {Lang}},\ }\href {https://doi.org/10.1007/978-3-642-01850-3} {\emph {\bibinfo {title} {{Quantum chromodynamics on the lattice}}}},\ Vol.\ \bibinfo {volume} {788}\ (\bibinfo  {publisher} {Springer},\ \bibinfo {address} {Berlin},\ \bibinfo {year} {2010})\BibitemShut {NoStop}%
\bibitem [{\citenamefont {Spanoudes}\ and\ \citenamefont {Panagopoulos}(2018)}]{Spanoudes2018}%
  \BibitemOpen
  \bibfield  {author} {\bibinfo {author} {\bibfnamefont {G.}~\bibnamefont {Spanoudes}}\ and\ \bibinfo {author} {\bibfnamefont {H.}~\bibnamefont {Panagopoulos}},\ }\bibfield  {title} {\bibinfo {title} {{Renormalization of Wilson-line operators in the presence of nonzero quark masses}},\ }\href {https://doi.org/10.1103/PhysRevD.98.014509} {\bibfield  {journal} {\bibinfo  {journal} {Phys. Rev. D}\ }\textbf {\bibinfo {volume} {98}},\ \bibinfo {pages} {014509} (\bibinfo {year} {2018})},\ \Eprint {https://arxiv.org/abs/1805.01164} {arXiv:1805.01164 [hep-lat]} \BibitemShut {NoStop}%
\bibitem [{\citenamefont {Spanoudes}\ \emph {et~al.}(2024)\citenamefont {Spanoudes}, \citenamefont {Constantinou},\ and\ \citenamefont {Panagopoulos}}]{Spanoudes:2024kpb}%
  \BibitemOpen
  \bibfield  {author} {\bibinfo {author} {\bibfnamefont {G.}~\bibnamefont {Spanoudes}}, \bibinfo {author} {\bibfnamefont {M.}~\bibnamefont {Constantinou}},\ and\ \bibinfo {author} {\bibfnamefont {H.}~\bibnamefont {Panagopoulos}},\ }\bibfield  {title} {\bibinfo {title} {{Renormalization of asymmetric staple-shaped Wilson-line operators in lattice and continuum perturbation theory}},\ }\href@noop {} {\bibfield  {journal} {\bibinfo  {journal} {arXiv e-prints}\ } (\bibinfo {year} {2024})},\ \Eprint {https://arxiv.org/abs/2401.01182} {arXiv:2401.01182 [hep-lat]} \BibitemShut {NoStop}%
\bibitem [{\citenamefont {Horsley}\ \emph {et~al.}(2004)\citenamefont {Horsley}, \citenamefont {Perlt}, \citenamefont {Rakow}, \citenamefont {Schierholz},\ and\ \citenamefont {Schiller}}]{Horsley2004}%
  \BibitemOpen
  \bibfield  {author} {\bibinfo {author} {\bibfnamefont {R.}~\bibnamefont {Horsley}}, \bibinfo {author} {\bibfnamefont {H.}~\bibnamefont {Perlt}}, \bibinfo {author} {\bibfnamefont {P.~E.~L.}\ \bibnamefont {Rakow}}, \bibinfo {author} {\bibfnamefont {G.}~\bibnamefont {Schierholz}},\ and\ \bibinfo {author} {\bibfnamefont {A.}~\bibnamefont {Schiller}} (\bibinfo {collaboration} {QCDSF}),\ }\bibfield  {title} {\bibinfo {title} {{One-loop renormalisation of quark bilinears for overlap fermions with improved gauge actions}},\ }\href {https://doi.org/10.1016/j.nuclphysb.2005.01.044} {\bibfield  {journal} {\bibinfo  {journal} {Nucl. Phys. B}\ }\textbf {\bibinfo {volume} {693}},\ \bibinfo {pages} {3} (\bibinfo {year} {2004})},\ \bibinfo {note} {[Erratum: Nucl.Phys.B 713, 601--606 (2005)]},\ \Eprint {https://arxiv.org/abs/hep-lat/0404007} {arXiv:hep-lat/0404007} \BibitemShut {NoStop}%
\bibitem [{\citenamefont {Costa}\ \emph {et~al.}(2021)\citenamefont {Costa}, \citenamefont {Karpasitis}, \citenamefont {Pafitis}, \citenamefont {Panagopoulos}, \citenamefont {Panagopoulos}, \citenamefont {Skouroupathis},\ and\ \citenamefont {Spanoudes}}]{Costa:2021iyv}%
  \BibitemOpen
  \bibfield  {author} {\bibinfo {author} {\bibfnamefont {M.}~\bibnamefont {Costa}}, \bibinfo {author} {\bibfnamefont {I.}~\bibnamefont {Karpasitis}}, \bibinfo {author} {\bibfnamefont {T.}~\bibnamefont {Pafitis}}, \bibinfo {author} {\bibfnamefont {G.}~\bibnamefont {Panagopoulos}}, \bibinfo {author} {\bibfnamefont {H.}~\bibnamefont {Panagopoulos}}, \bibinfo {author} {\bibfnamefont {A.}~\bibnamefont {Skouroupathis}},\ and\ \bibinfo {author} {\bibfnamefont {G.}~\bibnamefont {Spanoudes}},\ }\bibfield  {title} {\bibinfo {title} {{Gauge-invariant renormalization scheme in QCD: Application to fermion bilinears and the energy-momentum tensor}},\ }\href {https://doi.org/10.1103/PhysRevD.103.094509} {\bibfield  {journal} {\bibinfo  {journal} {Phys. Rev. D}\ }\textbf {\bibinfo {volume} {103}},\ \bibinfo {pages} {094509} (\bibinfo {year} {2021})},\ \Eprint {https://arxiv.org/abs/2102.00858} {arXiv:2102.00858 [hep-lat]} \BibitemShut {NoStop}%
\end{thebibliography}%

\end{document}